\documentclass[showkeys,printnumbers,amsmath,amssymb,12pt]{revtex4}    
\usepackage{graphicx}
\usepackage{dcolumn}
\usepackage{bm}
\usepackage{float}
\begin{document}

\begin{center}
\large{\textbf{Dynamic Studies of Scaffold-dependent Mating Pathway
in Yeast}}
\end{center}
\begin{center}
\textbf{Danying Shao$^{*\dag}$, Wen Zheng$^{*\dag}$, Wenjun
Qiu$^{*\dag}$,  Qi Ouyang $^{*\dag }$, and Chao Tang$^{*\dag\ddag}$}
\end{center}
\begin{center}
$^*$Center for Theoretical Biology and $^\dag$Department of Physics\\
Peking University,
 Beijing, 100871, China\\
\end{center}
\begin{center}
 $^\ddag$California
Institute for Quantitative Biomedical Research\\
 Departments of
Biopharmaceutical Sciences and Biochemistry and Biophysics\\
UCSF Box 2540, University of California at San Francisco, San
Francisco, CA 94143-2540\\
\end{center}
\begin{center}

Danying Shao and Wen Zheng contributed equally to this work.\\
Correspondence: Address reprint requests to Qi Ouyang (qi@pku.edu.cn) or Chao Tang (chao.tang@ucsf.edu).\\
\end{center}
\begin{center}
\textbf{Running Title:} Dynamics of yeast mating pathway\\
\end{center}
\begin{center}
\textbf{keywords:} Mating pathway, Scaffold, Shuttling, Phosphatase,
Concentration effect, MAPK Cascade
\end{center}

\newpage
\setlength{\baselineskip}{16pt} \textbf{Abstract}: The mating
pathway in \emph{Saccharomyces cerevisiae} is one of the best
understood signal transduction pathways in eukaryotes. It transmits
the mating signal from plasma membrane into the nucleus through the
G-protein coupled receptor and the mitogen-activated protein kinase
(MAPK) cascade. According to the current understandings of the
mating pathway, we construct a system of ordinary differential
equations to describe the process. Our model is consistent with a
wide range of experiments, indicating that it captures some main
characteristics of the signal transduction along the pathway.
Investigation with the model reveals that the shuttling of the
scaffold protein and the dephosphorylation of kinases involved in
the MAPK cascade cooperate to regulate the response upon pheromone
induction and to help preserving the fidelity of the mating
signaling. We explored factors affecting the dose-response curves of
this pathway and found that both negative feedback and
concentrations of the proteins involved in the MAPK cascade play
crucial role. Contrary to some other MAPK systems where signaling
sensitivity is being amplified successively along the cascade, here
the mating
 signal is transmitted through the cascade in an almost linear
 fashion.

\newpage
\begin{flushleft}
\textbf{INTRODUCTION}
\end{flushleft}

Cells have to respond to changes in the environment and/or to the
external stimuli. This is accomplished by signal transduction
pathways which sense the signal, transduce it and induce necessary
changes in the cell, e.g. in gene expression. One of the best
understood signaling pathways in eukaryotes is the mating pathway in
the budding yeast \emph{Saccharomyces cerevisiae} (1,2). Extensive
studies of the mating pheromone response have contributed much to
the understanding of the mechanisms of several conservative
biological modules (3), such as the G protein cycle (2,4) and the
mitogen-activated protein kinase (MAPK) cascade (2,5). Genetic,
biochemical and molecular analysis of the response have combined to
establish basic principles of the signaling and regulation. Many
important discoveries are made in the study of this pathway, for
instance, the concept of a kinase-scaffold protein (6) and the role
of regulator of G protein signaling (RGS) proteins in the pathway
(7).

The budding yeast can exist in either of the two types,
\emph{MAT}\textbf{a} or \emph{MAT}$\alpha$. These two types of cells
will mate when each one receives the mating signal, a peptide
pheromone, secreted by the opposite type (\textbf{a}-factor by
\emph{MAT}\textbf{a} and $\alpha$-factor by \emph{MAT}$\alpha$).
Once the pheromone binds to the seven-transmembrane-segment receptor
in the plasma membrane (Ste2 in \emph{MAT}\textbf{a} and Ste3 in
\emph{MAT}$\alpha$), the receptor is activated, which then activates
the heterotrimeric G protein that couples to it (Fig. 1). The
activated G protein transmits signal to multiple effectors,
resulting in the beginning of the MAPK cascade, which is embedded in
a scaffold protein Ste5. This cascade consists of three kinases:
Ste11 (MAPKKK), Ste7 (MAPKK), and Fus3 (MAPK). The activation of the
cascade finally leads to the phosphorylation of Fus3. The
phosphorylated MAPK then travels into the nucleus, and transmits the
signal to downstream effectors, leading to preparation for mating,
including the cell cycle arrest in G1 phase to assure synchronism of
the mating partners, the induction of new gene expression necessary
for mating, and the polarized growth in the direction of the
pheromone source.

Much qualitative and quantitative information in this pathway have
been documented. With the increasing amount of experimental data and
information, it is now possible to study this pathway quantitatively
at a systems level. Several mathematical models have been employed
to study this (8-12) and some other related systems (13-16), showing
that mathematical modeling and simulation can be a powerful method
in the analysis of functional and structural characteristics of
biological pathways.

We set up an Ordinary Differential Equation (ODE) model to describe
the mating pathway in budding yeast. Although several models have
been employed to illustrate mechanisms in the pathway, there has not
been one that integrates all the known essential features with a
comprehensive analysis of its dynamic properties. Some models were
constrained to a single step (8,12), while others oversimplified the
regulations and functions of the scaffold (10). In our model,
biochemical interactions, induced gene expressions which feed back
to the pathway, and translocations of key components such as the
scaffold protein Ste5 are all considered. Results from our model are
consistent with a wide range of experimental data. We then tested
the current understanding of regulations of cellular responses and
further explored the intrinsic mechanisms in the pathway, with
special interest in the role of the scaffold protein Ste5. We find
that the shuttling of the scaffold and dephosphorylation of the MAP
kinases cooperate to regulate the responses upon pheromone
induction, and to help keeping the fidelity of the mating pathway.
We further explored the mechanisms of the dose-response curves of
this pathway, and elucidated the role of enzyme concentration. We
found that instead of an ultrasensitive response as in some other
MAPK cascade (17), the mating
 signal here is transmitted through the cascade in an almost linear manner due to negative feedback.\\
\begin{flushleft}
\textbf{THE MODEL}
\end{flushleft}

We choose a mutant (TMY101), a \emph{MAT}\textbf{a} type of
\emph{Saccharomyces cerevisiae}, as the main modeling subject. In
this type of cells, the gene \emph{BAR1} is deleted. In a wild type
\emph{MAT}\textbf{a} cell, the product of \emph{BAR1} can be
excreted from the cells and cleave the $\alpha $-factor. In order to
simulate a continuous and constant $\alpha$-factor treatment, we use
this mutant in our model.

The mating response can be divided into three modules in a temporal
order: the activation of G protein cycle, the scaffold-depended MAPK
cascade, and the downstream effects of activated MAPK (See Fig. 1).
Viewing the response as a series of modules arranged in the temporal
order can help to better understand the signaling process. Yet in
our model, couplings and feedback between these modules are also
taken into account.

\textbf{\emph{The activation of G protein cycle}}. $\alpha $-factor
secreted by \emph{MAT}$\alpha$ binds to and hence activates the
seven-transmembrane-segment receptor (Ste2) on the plasma membrane
surface of \emph{MAT}\textbf{a}. Pheromone binding enhances
monoubiquitination of the receptor, and the ubiquitination in this
case serves as a signal for endocytosis and delivery to the vacuole
(18). This comprises a negative feedback loop (at short time
scales). In our model, this process is treated as a process of
accelerated degradation for simplicity. The synthesis of receptor
Ste2 is included;  the downstream effecter Ste12 is responsible for
the gene expression of Ste2. Thus, it comprises a positive feedback
(at long time scales).

The interaction between the activated receptor and G$\alpha$ leads
to some conformational changes, which enable G$\alpha $ to release
GDP and to bind GTP (19). G$\alpha\cdot$GTP can not interact with
G$\beta\gamma$, resulting in a release of G$\beta\gamma$ from the
receptor. The G$\gamma$ unit fixes the heterodimer on the plasma
membrane surface, while the G$\beta$ unit can interact with several
effectors to transmit the signal. In this sense, G$\alpha$ unit is a
negative regulator of the pathway; it plays a role in an
adaptational response to pheromone through preventing the
availability of G$\beta\gamma$ when there is no signal (20). The
G$\alpha\cdot$GTP can be hydrolyzed into G$\alpha\cdot$GDP, which
can re-associate with G$\beta\gamma$ into a heterotrimer. The cycle
of G protein is thus closed. Regulators of G protein signaling (RGS)
proteins can accelerate the hydrolyzation of G$\alpha\cdot$GTP (21).
In this pathway, the most important RGS protein is Sst2, which is
considered in our model. The gene expression of Sst2 is also
regulated by Ste12 in the downstream. Therefore, Sst2 is part of a
negative feedback loop that leads to the adaptation (8,22). Since
there is experimental evidence that the amount of G$\alpha$
increases significantly when the cells are treated with pheromone
(8), we add G protein synthesis in our model. It is commonly
accepted that Ste12 transcripts G$\alpha$ genes. According to the
above description, we formulate the reactions in G protein cycle as
follows: {\scriptsize
\[
\begin{gathered}
Ste2 + \alpha-factor\underset{{k2}}{\overset{{k1}}{\rightleftarrows}}Ste2_{active}  \hfill \\
  Ste2_{active} \xrightarrow{{k3}}      \hfill (\footnotesize{degradation})\\
  \xrightarrow[{k6}]{{k4,k5,ste12a}}Ste2    \hfill  (\footnotesize{synthesis}) \\
Ste2\xrightarrow{{k7}}\hfill\\
G\xrightarrow[{k8}]{{Ste2_{active} }}G\alpha\cdot GTP + G\beta \gamma \hfill \\
\xrightarrow[{k11}]{{k9,k10,ste12a}}G\xrightarrow{{k12}} \hfill \\
G\alpha\cdot GTP\xrightarrow[{Sst2_{active} ,k14}]{{k13}}G\alpha\cdot GDP \hfill \\
\end{gathered}\]
} {\scriptsize
\[
\begin{gathered}
  G\alpha\cdot GDP + G\beta \gamma \xrightarrow{{k15}}G \hfill \\
  \end{gathered}\]}\\
where k's are the kinetic parameters, the protein above or below the
arrow is enzyme or transcription factor of the reaction.

 \textbf{\emph{The scaffold depended MAPK pathway}}. The
released G$\beta\gamma$ has several effectors (23). One effector for
mating is Ste20, the first p21-activated protein kinase to be
identified in any eukaryote (24). Ste20 is also activated by Cdc42,
which is regulated by Cdc24. However, this process is not included
in our model because the Cdc42 binding domain of Ste20 has been
shown to be dispensable for pheromone signaling in yeast (25,26),
and there should be enough active Cdc24$^{GEF}$ and Cdc42
constitutively at the membrane to activate the amount of Ste20
required for initial signaling. Besides, mutants in Cdc24 do not
have much influence on the pathway (27,28).

Another effector of G$\beta\gamma$ is the scaffold protein Ste5. The
correlation between the disruption of the Ste4(G$\beta$)-Ste5
interaction and sterility confirms the importance of this
interaction in signal transduction (29). G$\beta\gamma$ can bind to
Ste5 on the LIM domain of Ste5, which is required for Ste11 (MAPKKK)
activation (30), probably through inducing a conformational change
that enhances Ste20-dependent activation of Ste11. Also it interacts
with Ste5 in the RING-H2 domain which is essential for Ste5
oligemerization (31).

Most scaffolds are contained in the nucleus during vegetative
growth. Upon pheromone induction they undergo enhanced exportation
from the nucleus and localize at the shmoo tip (1). Although the
detailed controlling mechanism of exportation of Ste5 is not clear,
it is plausible that mating pheromone increases the rate of Ste5
export (57). Here, we utilize an active control mechanism where the
import rate is kept constant, while the export rate is dependent on
the total concentration of the released G$\beta\gamma$. When there
is no signal, the export rate is very low, keeping most scaffolds in
the nucleus. When the mating signal opens the G-protein cycle,
released $G\beta\gamma$ enhances the export rate, driving scaffolds
to the shmoo tip. In this way, the localization of the scaffolds can
be regulated by G protein cycle.

The mating pathway is highly dependent on the scaffold protein Ste5.
First, Ste5 functions as an adapter protein. It recruits Ste11 to
the plasma membrane, where Ste20 is also tethered, to facilitate
Ste11's activation (28), triggering the MAPK cascade. Another
function for Ste5 is scaffolding. Ste5 tethers Ste11 (MAPKKK), Ste7
(MAPKK) and Fus3 (MAPK) to form a complex (32), keeping the kinases
and their substrates in proximity, as well as preventing the
influence of phosphatases. This function is supposed to be important
in enzyme regulation and in preventing cross talk (33,34).

When Ste5 is in the cytosol, it can form scaffold-kinase complexes
with Ste11, Ste7 and Fus3. Every kinase binding site on the scaffold
is in one of the three possible states: without a kinase, with an
unphosphorylated kinase, or with a dual phosphorylated kinase. So
for scaffold-kinase complexes in solution, there are
$3\times3\times3=27$ states: $B1, B2,...B27$ (See Fig. 2A).
G$\beta\gamma$ can bind to Ste20 and $Bi$ $   (i=1,2...27)$. Because
Ste20 is already on the plasmid membrane through the interaction
with Cdc42 before signaling, and scaffold must shuttle out from the
nucleus to bind to G$\beta\gamma$, we assume that G$\beta\gamma$
first binds to Ste20, then binds to $Bi$. Once $Bi$ binds to
G$\beta\gamma$Ste20 complex, it is fixed at the plasmid membrane and
the whole complex is denoted $Ci$ (see Fig. 2B). $Ci$ and $Bi$ are
the same in the interaction with MAPK kinases, except that $Ci$ can
phosphorylate Ste11 while $Bi$ can not.
 {\scriptsize
\[\begin{gathered}
 G\beta\gamma+Ste20\underset{{k19}}{\overset{{k18}}{\rightleftharpoons}}G\beta
\gamma Ste20 \hfill \\
Bi + G\beta \gamma Ste20\underset{{k17}}{\overset{{k16}}{\rightleftharpoons}}Ci    (i=1,2,...27)  \hfill \\
Ste5_{in}\underset{{k23}}{\overset{{k22}}{\rightleftharpoons}}B1 \hfill \\
\end{gathered}\]}\\
where "$Ste5_{in}$" denotes Ste5 in the nucleus, $B1$ denotes Ste5
outside of the nucleus, and k22 is dependent on the total
concentration of released G$\beta\gamma$:
$k22=0.0003+0.3\frac{G_{act}}{G_{act}+2500}, G_{act}=[G_\alpha\cdot
GTP]+[G_\alpha\cdot GDP]$.

Ste11 is the MAPKKK of yeast pheromone pathway, which consists of an
N-terminal regulatory domain and a C-terminal kinase region (1). The
interaction of these two domains keeps Ste11 in an anti-self state.
CBD domain in N-terminal contains serine and threonine residues that
can be phosphorylated by Ste20. Ste20-mediated phosphorylation of
these residues activates Ste11 (35). In addition to recruiting Ste11
to a pool of its activators Ste20, Ste5 also binds to Ste11 in its
N-terminal, making the CBD domain in N-terminal more accessible for
Ste20. Ste50 also helps to make the CBD domain more accessible to
Ste20 by a direct interaction between the SAM domain of itself and
the SAM domain of Ste11 (36), but is less essential than Ste5. Cells
lacking Ste50 are not truly sterile, thus we do not include Ste50 in
our model. Ste11pp phosphorylates the target residues in the
activation loops of Ste7 (MAPKK), and activates it (37). The
activated Ste7pp then phosphorylates, and activates its targets, the
MAPKs Fus3 and Kss1 (38). In the mating pathway, Fus3 plays a much
more important role, while Kss1 is the main MAPK in the
filamentation-invasion pathway in nitrogen starved cells (5). Thus
we do not consider Kss1 in our model, although analysis about
crosstalk will be given in the \emph{Discussion}. While Ste11 and
Ste7 are predominantly cytoplasmic proteins, Fus3 can shuttle
between the nucleus and the cytoplasm. It concentrates in the
nucleus after activation, thus bringing the signal to the nucleus
(7,35,39). There are also several feedback loops: Ste11 (MAPKKK)
undergoes ubiquitination and MAPK dependent degradation (40); Ste7
(MAPKK) is assumed to undergo enhanced degradation after
phosphorylation (41). In addition, Ste7pp in the scaffold is assumed
to be hyper-phosphorylated by activated Fus3pp, which reduces the
binding efficiency between Ste7pp and the scaffold sharply (42).

All the kinases can be categorized into two pools: on the scaffold
and in the solution. We assume that phosphorylation on scaffold
employs a processive mechanism, while phosphorylation in solution is
distributive(11). Processive mechanism means that the active kinase
collides with and binds to a substrate, phosphorylates it once, then
it may slides to align the second phosphorylation site of the
substrate with the active site of the kinase, phosphorylates the
substrate a second time before finally dissociates. Distributive
mechanism means that the active kinase collides with and binds to a
substrate, phosphorylates it once and releases the monophosphrylated
product, which then collides with a second molecule of the active
kinase, and is phosphorylated a second time (14). In our model, we
assume that a kinase is activated when and only when it is dual
phosphorylated, while a partial phosphorylated kinase possesses no
activity. Dual phosphorylation in a distributive manner could lead
to a sharp, sigmodial stimulus-response curve (14,17), leading an
all-or-none cell fate (14). However, the scaffold might diminish
this property if phosphorylations on the scaffold occur in a
processive manner (11). The dephosphorylations in the solution
employ the distributive mechanism while dephosphorylations in
scaffolds are precluded in our model due to sterical obstruction of
the phosphatases groups. The proteins responsible for
dephosphorylation of Ste11pp and Ste7pp are not clear. In the model,
we add two proteins with constant concentration to dephosphorylate
Ste11pp and Ste7pp respectively. There are several phosphatases for
Fus3pp: the dual-specificity phosphatase Msg5 (equally distributed
in nucleus and cytoplasm), and the tyrosine phosphatases Ptp3
(cytoplasm) and Ptp2 (nucleus), all of which can result in the
inactivation of Fus3pp (7,43,44). The basal level of Fus3
phosphorylation is controlled mainly by Ptp3, the amount of which is
constant during the stimulation (44). Pheromone treatment induces
the expression of Msg5 through the effects of Ste12 (43), which then
acts together with Ptp3 to inactivate Fus3pp. In our model, we use
MAPK-P with an initial concentration and with a synthesis rate
regulated by Ste12 to represent these three phosphatases. A recent
experiment shows that different inputs by Ste5 and Msg5 phosphatase
lead MAPK cascade to multiple outcomes (45), indicating that MAPK-P
is a key regulator in the network. Reactions of MAPK cascade in
cytosol are formulated as follows: {\scriptsize
 \[
 \begin{gathered}
  Ste11p + MAPKKK-P\underset{}{\overset{}{\rightleftharpoons}}Ste11pMAPKKK-P\xrightarrow{}Ste11 + MAPKKK-P\hfill \\
  Ste11pp + MAPKKK-P\underset{}{\overset{}{\rightleftharpoons}}Ste11ppMAPKKK-P\xrightarrow{}Ste11p + MAPKKK-P \hfill\\
  Ste7 + Ste11pp\underset{}{\overset{}{\rightleftharpoons}}Ste7Ste11pp\xrightarrow{}Ste7p + Ste11pp \hfill \\
  Ste7p + MAPKK-P\underset{}{\overset{}{\rightleftharpoons}}Ste7pMAPKK-P\xrightarrow{}Ste7 + MAPKK-P \hfill \\
  Ste7p + Ste11pp\underset{}{\overset{}{\rightleftharpoons}}Ste7pSte11pp\xrightarrow{}Ste7pp + Ste11pp \hfill \\
Ste7pp + MAPKK-P\underset{}{\overset{}{\rightleftharpoons}}Ste7ppMAPKK-P\xrightarrow{}Ste7p + MAPKK-P \hfill \\
Fus3_{out} + Ste7pp\underset{}{\overset{}{\rightleftharpoons}}Fus3_{out}Ste7pp\xrightarrow{}Fus3p_{out}+Ste7pp \hfill \\
Fus3p_{out} + MAPK-P_{out} \underset{}{\overset{}{\rightleftharpoons}}Fus3p_{out}MAPK-P_{out}\xrightarrow{}Fus3_{out} + MAPK-P_{out}  \hfill \\
  \end{gathered}
\]
\[
 \begin{gathered}
 Fus3p_{out} + Ste7pp\underset{}{\overset{}{\rightleftharpoons}}Fus3p_{out}Ste7pp\xrightarrow{}Fus3pp_{out} + Ste7pp  \hfill \\
Fus3pp_{out}  + MAPK-P_{out} \underset{}{\overset{}{\rightleftharpoons}}Fus3pp_{out}MAPK-P_{out}\xrightarrow{}Fus3p_{out} + MAPK-P_{out} \hfill \\
  Ste7 + Fus3_{out}\underset{{k25}}{\overset{{k24}}{\rightleftharpoons}}Ste7Fus3_{out} \hfill\\
Ste11\xrightarrow{{k26,Fus3pp_{out}}} \hfill \\
  Ste7pp\xrightarrow{{k27}} \hfill \\
  \end{gathered}
\]}\\
where p indicates once phosphorylation, and pp indicates twice
phosphorylation.

As for the scaffolds, we made the following assumptions in the
model: (a) Inactive kinase can bind to $Bi$ and $Ci$. On the
scaffold, this inactive kinase can either dissociate from the
scaffold without phosphorylation or undergo processive
phosphorylation before getting off the scaffold if its upstream
 kinase happens to be on the same scaffold and in the active state.
(b) Dephosphorylations on scaffolds are precluded due to sterical
obstruction. (c) There is no binding of partially activated kinases
to the scaffold proteins. For free fully activated kinases, only
Ste11pp can bind to the scaffold. Experimental evidence indicates
that active Ste5 can also accept Ste11pp activated by other pathways
and channel those signals to Fus3 (46), and that a greater amount of
scaffold proteins interact with Ste11 rather than the other two
kinases. As for Ste7, it undergoes hyper-phosphorylation by
activated Fus3pp, which accelerates its dissociation from the
scaffold (42). Fus3pp dissociates rapidly from the scaffold after
phosphorylation (47) and travels into the nucleus. Thus, the
reassociation of Ste7pp and Fus3pp to scaffold seems to be unlikely.
(d) Scaffold molecules do possess some catalytic properties (33), so
that the reaction rates within a scaffold complex are greater than
in the solution. Moreover, Ste7 and Fus3 can bind firmly (23), and
the residues for Ste7 binding in Fus3 are the same as the residues
for Ste5 binding (48), so it is reasonable to assume that Ste7
competes with Ste5 for binding to Fus3. Fig. 3 illustrates the
scaffold-dependent reactions of Ste11.

\textbf{\emph{Downstream effects}}: After activation, Fus3pp
dissociates rapidly from the scaffold, while the scaffold remains
tethered to the plasma membrane and partly in the solution (47,49),
acting as a platform for activation of many molecules of Fus3 and
leading to the propagation of the signal. The activated Fus3pp
transmits the signal into the nucleus, resulting in the activation
of transcription and the induction of cell cycle arrest. Fus3pp is
assumed to mediate the pheromone-induced transcription of
PRE-containing genes through phosphorylation and activation of at
least three nuclear proteins: Dig1, Dig2 and Ste12 (1,2). In
unstimulated cells, Dig1 and Dig2 bind to and thus repress Ste12
(50). Fus3pp phosphorylates Dig1, Dig2 and Ste12, and induces the
release of Ste12 from the complex (50,51). The free Ste12 then
interacts with other proteins of the transcription machinery and
thereby activates transcription of many different genes. Among the
products of these genes are proteins that activate (e.g. Fus3,
receptor) or inhibit (e.g. Msg5, Sst2) the pathway (2). Therefore,
the transcription affords several feedback loops in the pathway.
Another important substrate of Fus3pp is Far1. Activated Ste12
increases the transcription of Far1, and Fus3pp is able to
phosphorylate Far1 and thus to stabilize it (52). Far1 is a
bi-functional scaffold protein. In the cytoplasm, Far1 is involved
in polarized growth; while in the nucleus, it has a key function in
controlling cell cycle (53,54). Far1 inhibits Cln-Cdc28 complex, the
master regulator of the yeast cell cycle in G1 phase. In our model,
the binding of Far1 to Cln-Cdc28 is treated as a symbol for cell
cycle arrest. Ste12 is also the transcription factor of \emph{Bar1},
which is then excreted from the cell and inactivates $\alpha$-factor
(53). It is deleted in our model because most experiment results
that we use were from \emph{bar1$\triangle$} mutants. However,
\emph{bar1$\triangle$} could cause more than 100-fold sensitivity
increase in downstream transcription response (12). Reactions in the
downstream are:

{\scriptsize
\[
\begin{gathered}
\underset{{k30}}{\overset{{Fus3pp_{in} ,k28,k29}}{\rightleftharpoons}}Ste12_{active}\hfill\\
\xrightarrow[{}]{{Ste12_{active} ,k31,k5}}MAPK - P_{out}\hfill\\
  \xrightarrow[{}]{{Ste12a,k32,k5}}Fus3_{out} \hfill \\
  \underset{{k35}}{\overset{{Ste12_{active} ,k33,k34}}{\rightleftharpoons}}Far1\underset{{k37}}{\overset{{Fus3pp_{in} ,k36}}{\rightleftharpoons}}
  Far1pp_{in} \underset{{k39}}{\overset{{k38}}{\rightleftharpoons}}Far1pp_{out}  \hfill \\
  Far1pp_{out}  + G\beta \gamma \underset{{k41}}{\overset{{k40}}{\rightleftharpoons}}Far1pp_{out}G\beta \gamma  \hfill \\
   \end{gathered}
\]
\[
 \begin{gathered}Far1pp_{in}  + Cdc28\underset{{k43}}{\overset{{k42}}{\rightleftharpoons}}Far1pp_{in}Cdc28 \hfill \\
 \underset{{k46}}{\overset{{Ste12_{active} ,k44,k45}}{\rightleftharpoons}}Sst2_{active}  \hfill \\
Fus3_{in} \underset{{48}}{\overset{{k47}}{\rightleftharpoons}}Fus3_{out}\hfill\\
Fus3pp_{in} \underset{{k50}}{\overset{{k49}}{\rightleftharpoons}}Fus3pp_{out}\hfill \\
Fus3pp_{in}  + MAPK - P_{in} \underset{}{\overset{}{\rightleftharpoons}}Fus3pp_{in}MAPK - P_{in}\hfill\\
  \xrightarrow{}Fus3p_{in}  + MAPK - P_{in}  \hfill \\
  Fus3p_{in}  + MAPK - P_{in} \underset{}{\overset{}{\rightleftharpoons}}Fus3p_{in}MAPK - P_{in}\hfill\\
  \xrightarrow{}Fus3_{in}  + MAPK - P_{in}  \hfill \\
\end{gathered}
\]
}

 \textbf{\emph{Spatial location}}: we consider two compartments in
 the cell: the nucleus and the shmoo tip (a projection towards the direction of
pheromone formed as a result of polarized growth). The nucleus is
where downstream effects take place. The shmoo tip is where the many
signaling proteins are concentrated (55) and the main place for the
upstream reactions, including the G-protein cycle and Ste5-related
reactions. Thus in our model we neglected the rest part of the
cytosol. In other words, we "restricted" the cytosol to the shmoo
tip. The scaffold protein Ste5, the MAPK Fus3 (both activated and
inactivated), and Far1 are shuttling between the nucleus and the
shmoo tip. Ste5 is mostly sequestered in the nucleus in the absence
of pheromone while pheromone enhances nuclear exportation of Ste5
(56). Nuclear localization of Fus3 is slightly enhanced by the
pheromone treatment (7,39,57,58).

\textbf{\emph{The mathematical model}}: we employ a set of ordinary
differential equations (ODEs) to describe the changes in the
concentration of proteins involved in the mating pathway. Generally,
in a system of l biochemical species with the concentration $c_{i}$
(i=1,2,$\cdots$,l) and m biochemical reactions with the rates
$v_{j}$ (j=1,2,$\cdots$,m), the following series of equations can be
utilized to describe the biochemical mechanism in the system:
{\scriptsize
\[\displaystyle\frac{dc_{1}}{dt}=f_{1}(c_{1}, c_{2}, \cdots, c_{l})=n_{11}v_{1}+n_{12}v_{2}+\cdots+n_{1m}v_{m}\]
\[\displaystyle\frac{dc_{2}}{dt}=f_{2}(c_{1}, c_{2}, \cdots, c_{l})=n_{21}v_{1}+n_{22}v_{2}+\cdots+n_{2m}v_{m}\]
\[\vdots\]
\[\displaystyle\frac{dc_{l}}{dt}=f_{l}(c_{1}, c_{2}, \cdots, c_{l})=n_{l1}v_{1}+n_{l2}v_{2}+\cdots+n_{lm}v_{m}\]
}

The quantity $n_{ij}$ denotes the stoichiometric coefficient. The
rate of a reaction is a function of the concentrations of
substrates, products and probable effectors (10). If we treat the
gene expression as a special kind of reaction which can be described
with Hill functions, the equations listed above can be employed to
describe the dynamics of our system. In our model, all the unbound
substances in various phosphorylation states and complexes formed by
them are viewed as individual species. All complex formations,
dissociations, degradations, phosphorylations, and
dephosphorylations are treated as reactions. The parameters and
initial concentrations in the model are derived from experiments
whenever possible. For the remaining parameters, some are determined
by fitting the results of the model to indirect experiments; others
are estimated according to the mechanisms and similar reactions in
other organisms. The list of the model parameters as well as the
detailed ODEs are presented in \emph{SUPPLEMENTAL MATERIAL}. For
simulation, we use Matlab, version 6.5 (The MathWorks, Natick, MA).

\begin{flushleft}
\textbf{RESULTS}
\end{flushleft}
\begin{center}
{\textbf{\emph{A. Temporal characteristics}}}
\end{center}

\textbf{\emph{G-protein cycle}}. Fig. 4 summarizes the dynamics of
the G-protein cycle.  Upon saturated pheromone induction (1 $\mu$M
$\alpha$-factor), the level of the activated G-protein climbs up
rapidly, reaches its peak at about 30 s, and then gradually declines
to a bottom at about 7.5 minutes before it gradually increases
again, as shown in Fig. 4A. The simulation result (solid line) fits
quite well with the experiment data (12) (circles with error-bars).
One crucial factor that might contribute to enhancing the closure of
G-protein cycle is the endocytosis of activated receptor Ste2. This
hypotheses is supported by experiment with mutant
\emph{$Ste2^{300\triangle}$} (the C-terminal tail of the
$\alpha$-factor receptor gene STE2 is removed to impair its
endocytosis) (12). We slowed down the degradation rate of the active
Ste2 to simulate the \emph{$Ste2^{300\triangle}$} mutant. Consistent
with the experiment data, the closure of G-protein cycle is
apparently impaired and the amount of the activated G-protein levels
off after reaching its peak, as shown in Fig. 4B. The behavior of
\emph{$Ste2^{300\triangle}$} cells (dashed line for simulation, and
up-triangle with error-bars for experimental data) indicates that
endocytosis is a key factor that causes G protein cycle to close up.
Fig. 4A shows that after about t=10 minutes, the activated G-protein
continues to rise steadily. We attribute it to protein synthesis,
because that is the time scale for gene expression. To test this
hypothesis, we delete protein synthesis of all proteins considered
in our model, and find that the level of the activated G-protein
does not rise in the simulation. The behavior of cycloheximide
treated cells, as shown in Fig. 4B (dotted line for simulation and
square with error-bars for experiment data (12)), supports this
hypotheses. For comparison, time-course for TMY101 cells is also
shown in the figure (solid line for simulation, and circles with
error-bars for experiment data).

\textbf{\emph{Binding of Ste20 to G$\beta\gamma$}}: After
activation, G$\beta\gamma$ activates two effectors: Ste20 (MAPKKKK)
and $Bi$ (Scaffold in the solution), hence transmitting the signal
downwards. In our model, the time-course for the pheromone induced
binding of Ste20 to G$\beta\gamma$ fits well with experimental data
(24), (Fig. 4C). It shows that Ste20 binds quickly to G$\beta\gamma$
during the first 5 minutes. The binding slows down afterwards
 and then speeds up. This time-course seems
to follow the activation of G-protein cycle upstream (Fig. 4A),
consistent with the presumption that G$\beta\gamma$-dependent
activation rather than Cdc42-dependent activation of Ste20 is
critical in the mating pathway.

\textbf{\emph{Activation of MAPK pathway}}: Following the
recruitment of the scaffold protein Ste5 to the membrane, signal
passes down through the MAPK cascade (Fig. 4D). Note that except for
the gradual recruitment of Ste5, the signal transduction is very
fast.

\textbf{\emph{Downstream effects}}: Fig. 4E shows the activation of
Ste12, Far1ppG$\beta\gamma$ and Far1ppCdc28 to illustrate the
downstream effects of Fus3pp (MAPK). The formation of complex
Far1pp-Cln-Cdc28 is responsible for the cell cycle arrest, and
Far1pp-G$\beta\gamma$ causes the polarized growth and the formation
of the shmoo tip. Since G$\beta\gamma$ is a part of the upstream
complex involving Ste5 to provide a scaffold for the MAPK pathway,
excess G$\beta\gamma$ is not available until the pathway is
attenuated to some extent. Thus the curve for Far1-G$\beta\gamma$
begins to rise at 20 minutes after pheromone treatment, relatively
late compared to other downstream effectors.

\begin{center}
{\textbf{\emph{B. Features of the pathway}}}
\end{center}
\newcounter{mycounter}
\addtocounter{mycounter}{1}

\emph{\textbf{Scaffold shuttling and dephosphorylation cooperate to
regulate MAPK cascade quantitatively and to keep its fidelity to
mating signal}}: One of the most distinctive features of the mating
pathway is its dependence on scaffold. Interestingly, we found that
the amount of Ste5 localizing out of nuclear upon pheromone
induction coincides with activated Fus3pp (MAPK) in the
dose-response curve, as shown in Fig. 5A. Thus we speculate that the
mating pathway is tightly controlled by scaffold protein Ste5 and
its shuttling. It is highly possible that with different
concentrations of scaffolds out of nuclear, the efficiency of MAPK
cascade varies. When scaffold concentration is relative low, the
pathway efficiency should increase with scaffold concentration
because of the catalytical function and spacial protection function
of the scaffold. When scaffold concentration is too high, it may
reduce the mating efficiency because effective complex suitable for
signal transduction is hardly found (9). To investigate this
possibility, we shut down the shuttling of the scaffold protein Ste5
and varied the concentration of total Ste5 at the shmoo tip from 1nM
to 1000 nM. For the levels of pheromone induction, we varied the
concentration from 0.1 nM to 1000 nM. Consistent with our
expectation, the intensity of the output of MAPK cascade, indicated
by the amount of the activated Fus3pp, first increases with scaffold
concentration, then decrease, with the optimal scaffold
concentration around 100 nM, as shown in Fig. 5B. Thus, within our
simulation range, pathway efficiency monotonously increases with
scaffold available in the shmoo tip, and it is reasonable to suggest
that scaffold proteins have the ability to quantitatively control
the strength of signal transmission through the MAPK cascade. This
function of Ste5 could be achieved by facilitating Ste11's
activation by Ste20 through binding to G$\beta\gamma$, concentrating
MAPK cascade components, and sequestering them from inhibition by
phosphatases (See below).

Dephosphorylation is the other mechanism we speculate that might
contribute to controlling the activation of the mating pathway. To
test this hypothesis, we varied MAPK-P's concentration (both in the
nucleus and at the shmoo tip) from 100 nM down to 0 nM. As expected,
Fus3pp (MAPK) shows super-sensitivity upon pheromone induction, see
Fig. 5C. How does dephoshorylation control activation of MAPK
cascade? Since the kinases are exposed to phosphatase only when they
are in the cytosol and the scaffold could help to prevent the
influence of the phosphatase on the kinases bound to it, we suggest
that certain level of phosphatase concentration can keep the kinase
phosphorylation in the cytosol at a very low level, and thus
constrain the signal transduction on the scaffold. Hence, when the
phosphatases are attenuated, a large amount of activated Ste11pp,
Ste7pp, and Fus3pp could be accumulated in the cytosol even at lower
level of scaffold protein recruited to the shmoo tip, bypassing the
control of scaffold protein Ste5. In short, the shuttling of the
scaffold and the dephosphorylation of the MAP kinases cooperate to
control the activation of MAPK cascade quantitatively.

The cooperation between the scaffold shuttling and the MAPK
dephosphorylation is also crucial to the specificity of the pathway.
There are at least 5 MAPK signal transduction pathways in budding
yeast (5), some of which share the same proteins, such as Ste11 and
Ste7. A big puzzle is how specificity is achieved. To investigate
the mating pathway's ability to isolate inappropriate signals
leaking in from other pathways such as the filamentation-invasion
pathway, we tested the behavior of some mutants. We set all Ste11
molecules in the dual phosphorylated state at $t=0$ and shut off
MAPKKK-P to simulate the constitutive activation of Ste11pp in
invasive growth in the absence of mating pheromone. From the
time-course curve of Fig. 6A, we observed that although Ste7pp
(MAPKK) is activated to a relatively low extent, little Fus3pp
(MAPK) is stimulated. We then set all Ste7 molecules in the dual
phosphorylated state at $t=0$ and shut off MAPKK-P to simulate
constitutive Ste7 activation, and found that Fus3pp (MAPK) could
only be activated transiently in cytosol - the activation dropped
down immediately (within ten seconds) (inset in Fig. 6B). This
result is consistent with the experiments, revealing that persistent
activation by constitutive Ste7pp fails to support Fus3-dependent
mating in the yeast (42). Further simulation by removing MAPK-P
indicates that the later deactivation of Fus3pp is caused by
dephosphorylation (Fig. 6B). Note that only the kinases in the
solution are exposed to phosphatases, and the scaffold could help to
shield from the influence of phosphatases. Thus, phosphatases
constrain the signal on the scaffold. While on the scaffold, Ste7pp
prefers to phosphorylate Fus3 instead of Kss1, the signal
constrained on the scaffold will lead to the activation of Fus3,
rather than Kss1, the main MAPK in the filamentation-invasion
pathway in nitrogen-starved cells. That means that the activation of
Fus3 strictly relies on the scaffold protein. When there is no
pheromone induction, no scaffold protein is recruited to the shmoo
tip, and the phosphatases inhibit the basal activation of Fus3. With
pheromone treatment, the active scaffold proteins (the activation
means recruitment to the plasma membrane in our model) help to
assemble MAP Kinases cascade components and accelerate the
activation of Fus3pp, the main MAPK in the mating pathway. When
activation exceeds dephosphorylation, the signal will be transmitted
into the nuclear.

To conclude, when there is no pheromone induction, phosphatases
repress the activation of the mating pathway and prevent
inappropriate signals from leaking in. When pheromone exists,
scaffolds are recruited to the shmoo tip by the activated G protein,
gathering MAPK cascade components and sequestering them from
phosphatases so that the mating signal can be transmitted downwards.
Thus, mating pathway is highly depended on scaffolds. This
conclusion is consistent with experiments (45). Although there are
other factors that contribute to suppress the crosstalk between the
two pathways (34), the mechanism outlined above could also play an
important role.

\textbf{\emph{Desensitization to pheromone induction}}: The amount
of activated Fus3pp (MAPK) decreases with time even when the cells
are exposed to prolonged $\alpha$-factor induction, as shown in Fig.
7A  our wild type cell simulation (solid line). This indicates
desensitization effect. Desensitization is a key feature of the
pathway, which enables cells to reenter the cell cycle to resume
vegetative growth. We investigated the possible factors that might
contribute to this desensitization, and found that multiple negative
feedback loops, such as the degradation of Ste11 (MAPKKK) and Ste7pp
(MAPKK), the synthesis of Msg5 (MAPK phosphatase) and Sst2, should
be the major cause. As shown in Fig. 7A, a wild type cell with all
the negative feedback shows desensitization, but a mutant without
these negative feedback does not (dashed line): the activation of
Fus3pp does not decrease even after one hour treatment with
saturating pheromone.

Another important cause is the negative regulation of Ste7's (MAPKK)
binding ability to scaffold by Fus3pp (MAPK). In the scaffold,
Ste7pp, which undergoes feedback phosphorylation by activated
Fus3pp, dissociates more quickly from the scaffold (42), hence
exposes itself to ubiquitination and degradation (41,59). This
feedback can also accelerate the disassembling  of scaffold
complexes. In our model, we assume that the dissociation rate for
Ste7pp on scaffolds with both Ste7pp and Fus3pp is larger than that
for Ste7pp on scaffolds without Fus3pp. If we change former
parameter to be the same as the latter one, activation of Fus3pp
 continues to rise up after prolonged stimulation, as shown in
the dashed line in Fig. 7B. On the other hand, the dissociation rate
for Ste7pp from the scaffold without Fus3pp has to be relatively
slow to keep the intensity of the signal transduction. A mutant in
which the dissociation rate for the normal phosphorylated Ste7pp
(with no Fus3pp on the scaffold) is enhanced (dotted line in Fig.
7B) results in a very low intensity of the signal transduction. Thus
our model shows that differentiated binding abilities of Ste7 to the
scaffold ensure the right behavior of the MAPK cascade.

\textbf{\emph{Sensitivity to pheromone}}: Aside from the temporal
characteristics, the sensitivity to different levels of pheromone
induction is another key feature of the signal transduction pathway.
We compared the dose-response curves predicted by our model to those
observed in experiments (12,45), and there is a quantitative
agreement (Fig. 8). While our model specifically simulates the
\emph{$\Delta$Bar1} strain, our results are also consistent with
experiments of the wild type, taking into account the 100 fold
sensitivity shift (8,60). In our study, cells are assumed to be
treated with indicated concentration (0.001nM-1000 nM) of
$\alpha$-factor for 20 min. The response of every component to a
certain concentration of $\alpha$-factor is represented by the
maximum amount of that component. For the activated Ste2, the peak
is obtained within seconds; for G protein activation (measured by
the sum of G$\alpha$-GTP and G$\alpha$-GDP ), the maximum amount is
achieved at about 30 seconds; for Ste11pp, Ste7pp and Fus3pp, the
peak value appears at about 20 minutes. These different time scales
are consistent with experimental observations and illustrate the
characteristics of different components' activation. All these
curves are normalized. We fit the dose-response curves of Ste2a, G
protein activation, scaffold recruitment, Ste11pp, Ste7pp and Fus3pp
with the Hill Function:
\[[out]=\frac{A\times[in]^{n}}{[in]^{n}+th^{n}},\] where {\it n} is the Hill
coefficient and {\it th} the threshold where the response reaches
half its maximun (Table \ref{1}). Note that the sensitivity to
$\alpha$-factor is well conserved throughout the whole pathway, from
receptor Ste2 at the very beginning throughout the MAPK cascade.

As stated before, our model separates the whole mating pathway into
different modules. It is interesting to explore the dose-response
curve of each module. The first module is ligand binding. The reason
why the dose-response curve of the activated Ste2 is a Hill function
with $n\approx1$ is the reaction it takes. Consider the reaction
$Ste2+\alpha\underset{{k2}}{\overset{{k1}}{\rightleftharpoons}}Ste2_{act}$,
with the input concentration of $\alpha$ (on the left hand side)
fixed at [$\alpha$] and the total concentration of Ste2 fixed at
[Ste2]$_0$. At steady state:
\begin{equation}\label{Eq.X}
[Ste2_{act}]=\frac{A_{1}\times[\alpha]}{[\alpha]+th_1},
\end{equation}
with $A_1=[Ste2]_0$ and $th_1=k2/k1$, which takes the value
$th_1=5.0$ with our choice of parameters $k1$ and $k2$. The
coefficients derived from our simulation with our whole model are
$n=0.9$, $th=9.1$. The differences come from protein synthesis,
degradation and signal dependent feedback. When these effects are
deleted from the whole model, the simulation results agree perfectly
with the analysis.

We further explored the dose-response curve for G protein
activation. Again, for simplicity, we do not take into account in
the calculation the signal-dependent production, degradation and the
feedback of the RGS protein (we treat the amount of Sst2 as a
constant) and look for the steady state solution. The result of the
calculation is:
\begin{equation}\label{Eq.Y}
[G\beta\gamma]=\frac{[G]_0\times[Ste2_{act}]}{[Ste2_{act}]+\frac{k13'}{k8}},
\end{equation}
where $k13'=k13+k14[Sst2]$ . To get this result, we made the
assumption that the hydrolysis is a relatively slow process, so that
$\frac{k13'}{k15}<<[G]_{0}$. If we substitute Eq.(1) into Eq. (2),
we get:
\begin{equation}\label{Eq.Z}
[G\beta\gamma]=\frac{A_2\times[\alpha]}{[\alpha]+th_2},
\end{equation}
where $A_2=[G]_0\frac{A_1}{A_1+\frac{k13'}{k8}}$ and
$th_2=th_1\frac{k13'}{k8\times(A_1+\frac{k13'}{k8})}$. Eq. (3) is a
Hill function with $n=1.0$ and $th=2.9$. The simulation result of
our whole model is $n=1.0$ and $th=4.8$. Again, if the protein
synthesis, degradation and feedbacks are deleted, the simulation
results of $n=1.0$ and $th=2.9$ agree well with the analysis.

We see that the curve of $[G\beta\gamma]\sim[\alpha-factor]$ employs
the same type of function as the curve of
$[Ste2_{act}]\sim[\alpha$-factor], with the same Hill coefficient.
The only difference is that there is a shift in threshold.

The above analysis suggests that the amount of Sst2 can affect the
threshold of the dose-response curve. Fig. 9A indicates that Sst2,
which can accelerate the closure of the G protein cycle, is indeed a
key regulator of mating pathway's sensitivity. Comparison of the
dose-response curves for mutants SST2$\triangle$ (dotted line for
simulation, up-triangle with error-bars for experiment data), wild
type cells (solid line for simulation, squares for experiment data
(8)) and 2$\times$ SST2 (dashed line for simulation, down-triangle
for experiment data) clearly shows that the system is sensitive to
the amount of Sst2, which is consistent with the theoretical
analysis above. Another mutant we studied in this module is one with
excess G$\beta\gamma$ copies. Fig. 9B compares dose-response curves
for TMY101 cells (solid line for simulation, and squares for
experiment data) and cells with $2\times$ G$\beta\gamma$ (dashed
line for simulation, and up-triangles for experiment data(8)),
indicating that G$\beta\gamma$ alone is sufficient to switch on the
downstream signal transduction.

The next module in the pathway is the scaffold-dependent
phosphorylation cascade. Again, we do not take into account protein
synthesis and degradation, nor all the feedback. Here,
G$\beta\gamma$ is the input, while the concentration of Ste11pp,
Ste7pp and Fus3pp is chosen as the output. G$\beta\gamma$ first
binds to Ste20, then binds to scaffold in the solution to generate
$C$ (we use $C$ to indicate the ensemble of $C1$, $C2$,...$C27$)£¬
which then initiates MAPK cascade. We calculate the Hill
coefficients of Ste11pp, Ste7pp, and Fus3pp using G$\beta\gamma$ as
input, and find that each of them shows ultrasensitivity (Numerical
results: $n_{ste11pp}=2.3$, $n_{ste7pp}=2.2$, $n_{fus3pp}=1.8$).
Since G$\beta\gamma$ experiences two simple reversible reactions to
generate $C$, the relation between $C$ and G$\beta\gamma$ must be a
hill function with $n=1$. Therefore, the ultrasensitivity in this
module must arise from $C$ to Ste11pp, Ste7pp and Fus3pp. Let us
first consider the phosphorylation and dephosphorylation cycle of
Ste11. Ste11 can be dual phosphorylated on the scaffold in a
processive mechanism and dual dephosphorylated in the cytosol in a
distributive mechanism:
\[C+Ste11\underset{{off}}{\overset{{on}}{\rightleftarrows}}CSte11\underset{{}}{\overset{{p}}{\rightarrow}}C+Ste11pp.\]
\[Ste11pp+MAPKKK-P\underset{{d_{2}}}{\overset{{a_{2}}}{\rightleftarrows}}Ste11ppMAPKKK-P\underset{{}}{\overset{{p_{2}}}{\rightarrow}}Ste11p+MAPKKK-P\]
\[Ste11p+MAPKKK-P\underset{{d_{1}}}{\overset{{a_{1}}}{\rightleftarrows}}Ste11pMAPKKK-P\underset{{}}{\overset{{p_{1}}}{\rightarrow}}Ste11+MAPKKK-P\]
where we assume that Ste11pp leaves the scaffold so fast that
$CSte11pp$ can be neglected in the process. Introduce Michaelis
constant $K=\frac{off+p}{on}$, $K_{1}=\frac{d_{1}+p_{1}}{a_{1}}$,
$K_{2}=\frac{d_{2}+p_{2}}{a_{2}}$, and define
$\alpha_{1}=\frac{pK_{1}}{p_{1}K}$,
$\alpha_{2}=\frac{pK_{2}}{p_{2}K}$,
$p'=\frac{p_{1}p_{2}}{p_{1}+p_{2}}$,
$K'=\frac{p_{1}K_{2}+p_{2}K_{1}}{p_{1}+p_{2}}$, we get:
\begin{equation}\label{Eq.W}
\left[ {Ste11pp } \right] = \frac{{\alpha _2 }}{{\alpha _1  + \alpha
_2 }}G\left( {pC_{total} ,p'[MAPKKK-P]_0
,\frac{K}{{[Ste11]_0}},\frac{{K'}}{{[Ste11]_0 }}} \right)[Ste11]_0,
\end{equation}
where\[ G\left( {u,v,M,N} \right) = \frac{{2uN}}{{v - u + vM + uN +
\sqrt {\left( {v - u + vM + uN} \right)^2  - 4\left( {v - u}
\right)uN} }},
\]
\[u=p[Ste11]_0,~~~~v=p'[MAPKKK-P]_0,~~~~~~M=\frac{K}{[Ste11]_0},~~~~~~N=\frac{K'}{[Ste11]_0}.\]
The above G function can be fitted to a sigmoidal curve (Hill
function)(61), with the Hill coefficient and the threshold value to
be

\begin{equation}\label{Eq.U}
n=\frac{1}{\log_{81}{\frac{81(M+0.1)(N+0.1)}{(M+0.9)(N+0.9)}}},
\end{equation}
\begin{equation}\label{Eq.U}
th=\frac{p'(1+2M)}{p(1+2N)}[MAPKKK-P]_0.
\end{equation}

According to the parameters in the Ste11 cycle,
$M=K/[Ste11]_0\approx 0.125$ and $N=K'/[Ste11]_0=0.255$. So Eq. (7)
approximates a Hill function with $n\approx2.6$ and $th\approx 21
nM$. Therefore, it is the zeroth-order ultrasensitivity that leads
to the ultrasensitive response of Ste11pp. Similarly, in the Ste7
cycle, $M\approx 0.183$ and $N=0.324$; in the Fus3 cycle, $M\approx
0.135$ and $N=0.181$. Therefore, Ste7 and Fus3 are also located in
the zeroth-order region, making them ultrasensitive.

The analysis above together with Eq. (6) indicates that the relation
between Ste11pp, Ste7pp, Fus3pp and $\alpha$ should exhibit
ultrasensitivity (numerical result: $n_{ste11pp}=2.1$,
$n_{ste7pp}=1.8$, $n_{fus3pp}=1.9$). This appears to contradict the
results in Table \ref{1} where all the dose-response curves overlap
with Hill coefficient $n\approx1$. The reason for this contradiction
is that in the analysis we cut off all the feedback in the original
model. There are all together nine feedback in the whole model, six
of which are negative feedback: the transcription of Sst2, MAPK-P,
the degradation of activated Ste2, Ste11, Ste7pp, and the
hyper-phosphorylation of Ste7pp. When $[\alpha]$ is low, all these
negative feedback are kept low, leaving the output nearly
unaffected. When $[\alpha]$ is high, the negative feedback will also
be strong, significantly reducing the output. Therefore, negative
feedback can make the dose-response curve less steep. It may be the
counterbalance between the zeroth-order ultrasensitivity and the
negative feedback that keeps the Hill coefficients of Ste11pp,
Ste7pp, and Fus3pp remain 1. In order to test this view, we add each
of the six negative feedback into the simplified model where all
feedback are cut off (Table \ref{2}. Numerical result shows that
each feedback can reduce the Hill coefficients, and if all the
negative feedback are added together, the Hill coefficients can
decrease to nearly 1, indicating that negative feedback indeed can
reduce the Hill coefficients.

We further study how the Hill coefficients of Ste11pp, Ste7pp,
Fus3pp depend on the scaffold, the substrate and the phosphatase
concentrations (Table \ref{3}. First, we increase Ste11, Ste7 or
Fus3 concentration by 10 times, and find that the Hill coefficients
only change a little. The probable reason is that the original
kinase concentration is already much larger than the scaffold
concentration, so when Ste11, Ste7 or Fus3 concentration increases,
the added part could not get to the scaffold to be phosphorylated,
and thus does not contribute to response. Then, we increase kinase
concentration and scaffold concentration together, and find that
Hill coefficients have a substantial increase.
 This is because when scaffold and kinase concentration increase together,
effective substrate concentration increases. Therefore, the
zeroth-order ultrasensitivity becomes more significant, while
feedback, which is mainly dependent on the downstream regulation,
does not increase as fast. Table \ref{3} also shows that when each
phosphatase concentration is decreased by 10 times, Hill
coefficients arise as well. This is because low phosphatase
concentration allow more substrate to be phosphorylated in the
cytosol in a distributive mechanism, which is regarded as another
mechanism to generate ultrasensitivity aside from the zeroth-order
ultrasensitivity (17). Thus, the scaffold, the substrate and the
phosphatase concentration play an important role in determining the
Hill coefficients of the MAPK pathway.

\textbf{\emph{Analysis of parameters}}: Due to the lack of
experimental data to determine all the parameters, it is necessary
to analyze the sensitivity of the system to changes of the
parameters. To do this, we define a quantity:
\[D=\sqrt{\overline{(\frac{([Fus3pp]_{cal}-[Fus3pp]_{ori}}{[Fus3pp]_{ori}})^{2}}}\]
where $[Fus3pp]_{cal}$ denotes the calculated output when a
parameter is changed and $[Fus3pp]_{obs}$ denotes the original
output, and the bar denotes the average of the relative variance of
the output over an input ($\alpha$-factor) range of $10^{-3}$ nM to
$10^3$ nM. We multiply and divide one parameter by 2 at each time,
calculate D, and then take the average for the two situations of
increasing and decreasing the parameter. The most sensitive
parameters are listed in Table \ref{4}. Note that these parameters
all have direct or indirect experimental support, which is
reassuring. These parameters also give some clues about the
mechanisms of the mating pathway. The parameters which influence the
outcome of the pathway most are those involved in the receptor
activation, which is consistent with our finding that the shape of
dose-response curve is determined by the first step in the
pathway--the receptor activation. The parameter in the production of
receptor Ste2 is also essential to the outcome of the system. The
other two influential parameters are the rates of the kinases to get
off the scaffold, so the scaffold protein is also an important
factor in this pathway.
\begin{flushleft}
\textbf{DISCUSSION}
\end{flushleft}

Our model describes the whole mating pathway comprising of the G
protein cycle, the scaffold-dependent MAPK cascade and the
downstream effects in the nucleus. We have investigated multiple
features of the pathway, including its various characteristic times
scales, desensitization, scaffold's effect, specificity, sensitivity
to different levels of pheromone induction, the role of feedback,
and sensitivity amplification. Although many the parameters in our
model do not have solid experimental support and the detailed
mechanisms of some steps are still not clear, the results given by
our model are consistent with the current understanding of the
pathway and with a wide range of experimental data.

The duration and sensitivity of the mating pathway have to be
tightly regulated; an inappropriate activation of Fus3 will block
the normal invasive growth. Our model shows that activation of the
mating pathway attenuates with time even when the pathway is exposed
to prolonged pheromone induction. This desensitization is attributed
to several feedback such as the enhanced degradation of Ste7
(MAPKK). These feedback enable the cell to recover from mating and
continue their vegetative growth upon prolonged pheromone induction.

Evidence shows that oligomerization of the scaffold protein is
required for its activation (31,57,62,63). However, oligomerization
is not included in our model for simplicity. Since we consider all
possible complexes involving the scaffold protein, taking the
oligomorization into account would make the model extremely
complicated (with 27$\times$27=729 possible complexes involving the
scaffold protein). Moreover, there are few experimental data
concerning the interactions between the scaffold protein and the
kinases. Therefore, we have to make some simplifications. A clue for
the simplification is the experimental evidence that nuclear export
and shmoo tip recruitment of Ste5 are coordinated with
oligomarization (57,61). Thus we use nuclear export as a controlling
step. In our model, the recruitment of the scaffold protein to the
shmoo tip implies its activation, including the effect of nuclear
transportation and oligomorization. However, oligomarization may
have other effect in the pathway besides the activation of scaffold
protein. Due to the lack of experimental data in this process, it is
difficult to consider it in detail in our model. Experimental
progress about this process is much needed to further improve the
model.

The scaffold protein undergoes continuous shuttling and enhanced
exportation upon pheromone induction. The obvious question is why
yeast cells take so much trouble to shuttle a huge protein through
the nucleus when it functions predominantly in cytoplasm. Our model
indicates that nuclear shuttling might be a key step controlling the
availability of the scaffold protein to the pathway. Since the
activation of the MAPK cascade in the mating pathway is dependent on
the scaffold protein, whether and how many scaffold proteins are
available determine whether and how efficient the MAPK pathway is
stimulated. However, the mechanism of scaffold shuttling are still
not clear. The complex Msn5p/Ste21 is suggested to be responsible
for the export of Ste5p. Further work is required to establish the
accurate and detailed mechanism of this controlling step. In our
model, we employ an active control mechanism. That is, the scaffold
export rate is dependent on the concentration of the separated
G$\beta\gamma$, which is released by the phemorone. The more the
pheromone, the more the activated G$\beta\gamma$, and the higher the
export rate. In other words, G protein cycle controls the shuttling
of the scaffold and the concentration of scaffold at the shmoo tip.

Another function of the scaffold is its role in keeping the
pathway's fidelity to the signal. The capability of the scaffold
protein to prevent kinases from dephosphorylation assures the mating
pathway's dependence on
 the scaffold protein, and the availability of scaffold in the shmoo tip is further controlled
by the G protein cycle. Recently the specificity of different
pathways in yeast is under intensive study (34). It is interesting
to see that different cellular signals, which can be transmitted by
the same components, result in distinct responses. Especially, the
haploid invasive growth pathway employs the same MAPK components
(Ste11 as MAPKKK, Ste7 as MAPKK, Fus3 and Kss1 as MAPK) as the
mating pathway, except that Fus3 is more active during mating while
Kss1 is preferentially activated during invasive growth. Then how
are the different outputs controlled? Our model suggests that
dephosphorylation and scaffolds work in coordination to prevent
improper signal from leaking in, and thus contributing to the mating
pathway's fidelity to pheromone induction. Further work is needed to
include the parallel pathway of the invasive growth, through which a
more comprehensive understanding of specificity might be obtained.

MAPK cascade which is conserved in all eukaryotic cells is composed
of both phosphorylation and dephosphorylation. Our model reveals
that dephosphorylation has several roles in the mating pathway. It
is obvious that it contributes to the desensitization of the pathway
which enables the cell to reenter the cell cycle. Furthermore, it
cooperates with the shuttling of the scaffold proteins, to realize
other important features of the signaling pathway. First, it helps
to preserve the consistence of the sensitivity from the G-protein
cycle to the MAPK cascade. Second, the amount of MAPK-P, together
with the scaffold protein, contributes to the pathway fidelity.

Notably, we suggest that negative feedback plays an important role
in the experimentally observed preservation of sensitivity along the
MAPK cascade. Whether or not the sensitivity is amplified as the
MAPK cascade descends is also determined by the concentrations of
the kinases and phosphatases involved. We call for new experiments
to test this
hypothesis.\\

\begin{flushleft}
\textbf{ACKNOWLEDGEMENT}
\end{flushleft}

We thank members of the Peking University Center for Theoretical
Biology for their valuable suggestions. This research is supported
by National Key Basic Research Project of China, Chinese Natural
Science foundation, and the Chun-Tsung endowment at Peking
University. C.T. acknowledges support from the Sandler Family Supporting Foundation.\\
\begin{flushleft}
\textbf{REFERENCES}\\
1. Bardwell, L. 2004. A walk-through of the yeast mating pheromone
response pathway. \emph{Peptides.} 25:1465-1476.\\
 2. Dohlman, H.
G., and J. W. Thorner. 2001. Regulation of G protein-initiated
signal transduction in yeast: paradigms and principles. \emph{Annu.
Rev. Biolchem.} 70:703-754. \\
3. Wang, Y., and H. G. Dohlman. 2004. Pheromone signaling mechanisms
in yeast: a prototypical sex machine. \emph{Science.} 306:1508-1509.
\\
4. Dohlman, H. G. 2002. G protein and pheromone signaling.
\emph{Annu. Rev. Physiol.} 64:129-152. \\
5. Gustin, M. C., J. Albertyn, M. Alexander, and K. Davenport. 1998.
MAP kinase pathways in the yeast \emph{Saccharomyces cerevisiae}.
\emph{Microbiol. Mol. Biol. Rev.} 62:1264-1300. \\
6. Choi, K., B. Satterberg, D. M. Lyons, and E. A. Eloin. 1994. Ste5
tethers multiple protein kinases in the MAp kinase cascade required
for matin in \emph{S. cerevisiae}. \emph{Cell.} 78:499-512. \\
7. Blackwell, E., I. M. Halatek, H. N. Kim, A. T. Ellicott, A. A.
Obukhov, and D. E. Stone. 2003. Effect of the pheromone-responsive
G$\alpha$ and phosphatase proteins of \emph{saccharomyces
cerevisiae} on the subcellular localization of the Fus3
Mitogen-Activated Protein Kinase. \emph{Mol. Cell.
Biol.}23:1135-1150. \\
8. Hao, N., N. Yildirim, Y. Wang, T. C. Elston, and H. G. Dohlman.
2003. Regulators of G protein signaling and transient activation of
signaling. \emph{J. Biol. chem.} 278:46506-46515. \\
9. Kholodenko, B. N.. 2000. Negative feedback and ultrasensitivity
can bring about oscillation in the mitogen-activated protein kinase
cascades. \emph{Eur.J. Biochem.} 267:1583-1588. \\
10. Kofahl, B., and E. Klipp. 2004. Modeling the dynamics of the
yeast pheromone pathway. \emph{Yeast.} 21:831-850. \\
11. Levchenko, A., J. Bruck, and P. W. Sternberg. 2000. Scaffold
proteins may biphasically affect the levels of Mitogen-activates
protein signaling and reduce its threshold properties. \emph{PNAS.}
97:5818-5823. \\
12. Yi,T., H. Kitano, and M. Simon. 2003. A quantitative
characterization of the yeast heterotrimeric G protein
cycle. \emph{PNAS.} 100:10764-10769. \\
13. Bhalla, U. S., P. T. Ram, and R. Lyengar. 2002. MAP kinase
phosphatase as a locus of flexibility in a mitogen-activated protein
kinase signaling network. \emph{Science Reports.} 297:1018-1023.
\\
14. Ferrell, J. E., and R. R. Bhatt. 1997. Mechanistic studies of
the dual phosphorylation of mitogen-activated protein kinase.
\emph{J.
Biol. Chem.} 272:19008-19016. \\
15. Ferrell, J. E., and E. M. Machleder. 1998. The biochemical basis
of an all-or-none cell fate ewith in \emph{Xenopus Oocytes}.
\emph{Science Reports.}
280:895-898. \\
16. Xiong, W., and E. Ferrell. 2003. A positive-feedback-based
bistable 'memory module' that governs a cell fate decision.
\emph{Nature.} 426:460-465.\\
17. Huang, C. F., and J. E. Ferrell. 1996. Ultrasensitivity in the
mitogen-activated protein kinase cascade. \emph{Proc. Natl. Acad.
Sci. USA.} 93:10078-10083. \\
18. Hicke, L., and H. Reizman. 1996. Ubiquitination of a yeast
plasma membrane receptor signals its ligand-stimulated endocytosis.
\emph{Cell.} 84:277-288. \\
19. Conklin, B. R., and H. R. Bourne. 1993. Structural elements of
G$\alpha$ subunits that interact with G$\beta\gamma$, receptors and
effectors. \emph{Cell.} 73:631-641.\\
20. Cole, G. M., D. E. Stone, and S. I. Reed. 1990. Stoichiometry of
G protein subunits affects the \emph{Saccharomyces cerevisiae}
mating pheromone signal transduction pathway. \emph{Mol. Cell.
Biol.} 10:510-517. \\
21. Dohlman, H. G., and J. Thorner. 1997. RGS proteins and signaling
by heterotrimeric G protein. \emph{J. Biol. Chem.} 272:3871-3874.
\\
22. Dohlman, H. G., D. Apaniesk, Y. Chen, J. Song, D. Nusskern.
1995. Inhibition of G-protein signaling by dominant gain-of-function
mutations in Sst2p, a pheromone desensitization factor in
\emph{Saccharomyces cerevisiae}. \emph{Mol. Cell. Biol.}
15:3635-3643. \\
23. Bardwell, L., J. G, Cook, E. C. Chang, B. R. Cairns, and
J.Thorner. 1996. Signaling in the yeast pheromone response pathway:
specific and high-affinity interaction of the mitogen-activated
protein (MAP) kinases Kss1 and Fus3 with the upstream MAP kinase
kinase Ste7. \emph{Mol. Cell. Biol.}16:3637-3650. \\
24. Leeuw, T., C. Wu, J.D. Schrag, M. Whiteway, D. Y. Thomas, and E.
Leberer. 1998. Interaction of a G-protein $\beta$-subunit with a
conserved sequence in Ste20/PAK family protein kinases.
\emph{Nature.} 391:191-195. \\
25. Leberer, D.. 1997. Functional characterization of the Cdc42p
binding domain of yeast Ste20p protein kinase. \emph{EMBO J}
16:83-97. \\
26. Peter, M., A. M. Neiman, H. O. Park, M. Van Lohuizen, and I.
Herskowitz. 1996. Functional analysis of the interaction between the
small GTP binding Cdc42 and the Ste20 protein kinase in
yeast. \emph{EMBO J.} 15:7046-7059. \\
27. Lamson, R. E., M. J. Winters, and P. M. Pryciak. 2002. Cdc42
regulation of kinase activity and signaling by the yeast
p21-activates kinase Ste20. \emph{Mol. Cell. Biol.} 22:2939-2951.
\\
28. Pryciak, P. M., and F. A. Huntress. 1998. Membrane recruitment
of the kinase cascade scaffold protein Ste5 by the G$\beta$$\gamma$
complex underlies activation of the yeast pheromone response
pathway. \emph{Genes Dev.} 12:2684-2697. \\
29. Dowell, S. J., A. L. Bishop, S. L. Dyos, A. J. Brown, and M. S.
Whiteway. 1998. Mapping of a yeast G $\beta$$\gamma$ signaling
interaction. \emph{Genetics.} 150:1407-1417. \\
30. Feng, Y., L. Y. Song, E. Kincaid, S. K. Mahanty, and E. A.
Elion. 1998. Functional binding between G$\beta$ and LIM domain of
Ste5 is required to activate the MEKK Ste11. \emph{Curr. biol.}
8:267-278. \\
31. Sette, C., C. J. Inoyue, S. L. Stroschein, P. J. Iaquinta, and
J. Thorner. 2000. Mutational analysis suggests that activation of
the yeast pheromone response mitogen-activated protein kinase
pathway involves conformational changes in the Ste5 scaffold
protein. \emph{Mol. Biol. Cell.} 11:4033-4049. \\
32. Printen, J. A., and G. F. Sprague. 1994. Protein-protein
interaction in the yeast pheromone response pathway: Ste5 interacts
with all members of the MAP kinase cascade. \emph{Genetics.}
138:609-619. \\
33. Park, S., A. Zarrinpar, and W. A. Lim. 2003. Rewiring MAP kinase
pathways using alternative scaffold assembly
mechanisms. \emph{Science.} 299:1061-1064. \\
34. Schwartz, M.A., and H.D.Madhani. 2004. Principles of MAP kinase
signaling specificity in \emph{Saccharomyces Cerevisiae}.
\emph{Annu.Rev.Genet.} 38:725-48.\\
 35. Van Drogen, F., S. M.
O'Rourke, V. M. Stucke, M. Jaquenoud, A. M. Neiman, and M. Peter.
2000. Phosphorylation of the MEKK Ste11p by the PAK-like kinase
Ste20p is required for MAP kinase signaling in vivo. \emph{Curr.
Biol.}10:630-639. \\
36. Kwan, J. J., N. warner, T. Pawson, and L. W. Donaldson. 2004.
The solution strcture of the \emph{S. cerevisiae} Ste11 MAPKKK SAM
domain and its partnership with Ste50. \emph{J. Mol. Biol.}
342:681-693. \\
37. Neiman, A. M., and I. Herskowitz. 1994. Reconstitution of a
yeast protein kinase cascade in vitro: Activation of the yeast MEK
homolog STE7 by STE11. \emph{Proc. Natl. Acad. Sci. USA.}
91:3398-4402. \\
38. Errede, B., A. Gartner, Z. ZHou, K. Nasmyth, and G. Ammerer.
1993. MAP kinase-related Fus3 from \emph{S. cerevisiae}
is activated by Ste7 in vitro. \emph{Nature.} 362:261-264. \\
39. Choi, K., J. E. Kranz, S. K. Mahanty, K. Park, and E. A. Elion.
1999. Characterization of Fus3 localization: active Fus3 localizes
in complexes of varying size and specific activity. \emph{Mol. Bio.
Cell.} 10:1553-1568. \\
40. Esch, R. K., and B. Errede. 2002. Pheromone induction promotes
Ste11 degradation through a MAPK feedback and ubiquitin-dependent
mechanism. \emph{PNAS.} 99:9160-9165. \\
41. Wang, Y., and H. G. Dohlman. 2002. Pheromone-dependent
ubiquitination of the mitogen-activated protein kinase kinase Ste7.
\emph{J. Biol. Chem.}
277:15766-15772. \\
42. Maleri, S., Q. Ge, E. A. Hackett, Y. Wang, H. G. Dohlman, and B.
Errede. 2004. Persistent activation by constitutive Ste7 promotes
kss1-mediated invasive growth but fails to support Fus3-dependent
mating in yeast. \emph{Mol. Cell. Biol.}
24:9221-9238. \\
43. Doi, K., A. Gartner, G. Ammerer, B. Errede, H. Shinkawa, K.
Sugimoto. 1994. MSG5, a novel protein phosphatase promotes adaption
to pheromone response in \emph{S. cerevisiae}. \emph{EMBO
J.} 13:61-70. \\
44. Zhan, X. L., R. J. Deschenes, and K. L. Guan. 1997. Differential
regulation of FUS3 MAP kinase by tyrosine-specific phosphatases
PTP2/PTP3 and dual-specificity phosphatase MSG5 in
\emph{Saccharomyces cerevisiae}. \emph{Gene.
Dev.} 11:1690-1702. \\
45. Andersson, J., D. M. Simpson, M. Qi, Y. Wang, and E. A. Elion.
2004. Differential input by Ste5 scaffold and Msg5 phosphatase route
a MAPK cascade to multiple outcomes. \emph{J. EMBO.} 23:2564-2576.
\\
46. Flatauer, L. J., S. F. Zadeh, and L. Bardwell. 2005.
Mitogen-Activated Protein Kinase with distinct requirements for Ste5
scaffolding influence signaling specificity in \emph{Saccharomyces
cerevisiae}. \emph{Mol. Cell. Biol.} 25;1793-1803. \\
47. Van Drogen, F., and M. Peter. 2001. MAP kinase dynamics in
yeast. \emph{Biol. Cell.} 93:63-70. \\
48. Kusari, A. B., D. M. Molina, W. Sabbagh, C. S. Lau, and L.
Bardwell. 2004. A conserved protein interaction network involving
the yeast MAP kinases Fus3 and Kss1.
\emph{J. Cell. Biol.} 164:267-277. \\
49. Van Drogen, F., V. M. Stucke, G. Jorritsma and M. Peter. 2001.
MAP kinase dynamics in response to pheromone in budding yeast.
\emph{Nature Cell Biol.} 3:1051-1059. \\
50. Tedford, K., S. Kim, D. Sa, K. Stevens, and M. Tyers. 1997.
Regulation of the mating pheromone and invasive growth responses in
yeast by two MAP kinase substrates. \emph{Curr. Biol.} 7:228-238.
\\
51. Cook, J. G., L. Bardwell, S. J. Kron, and J. Thorner. 1996. Two
novel targets of the MAP kinase Kss1 are negative regulators of
invasive growth in the yeast \emph{Saccharomyces cerevisiae}.
\emph{Genes. Dev.} 10:2831-2848.\\
52. Gartner, A., A. Jovanovic, D. I. Jeoung, S. Bourlat, F. R.
Cross, and G. Ammerer. 1998. Pheromone-dependent G1 cell cycle
arrest requires Far1 phosphorylation, but may not involve inhibition
of Cdc28-Cln2 kinase, in vivo. \emph{Mol. Cell. Biol.} 18:3681-3691.
\\
53. Ballensiefen, W., and H. D. Schmitt. 1997. Periplasmic
\emph{Bar1} protease of \emph{Saccharomyces cerevisiae} is active
before reaching its extracellular destination. \emph{Eur. J.
Biochem.}
247:142-147. \\
54. Chang, F., and I. Herskowitz. 1990. Identification of a gene
necessary for cell cycle arrest by a negative growth factor of
yeast: FAR1 is an inhibitor of a Gly cyclin, CLN2. \emph{Cell.}
63:999-1011. \\
55. Bagnat, M., and K. Simon. 2002. Cell surface polarization during
yeast mating. \emph{Proc. Natl. Acad. Sci. USA.} 99:14183-14188.
\\
56. Mahanty, S. K., Y. Wang, F. W. Farley, and E. A. Elion. 1999.
Nuclear shuttling of yeast scaffold Ste5 is required for its
recruitment to the plasma membrane and activation of the mating MAPK
cascade. \emph{Cell.} 98:501-512. \\
57. Elion, E. A.. 2001. The Ste5p
scaffold. \emph{J. Cell. Sci.} 114:3967-3978. \\
58. Elion, E. A.. 2000. Pheromone response, mating and cell biology.
\emph{Current opinion in cell. biol.} 3:573-581. \\
59. Wang, Y., Q. Ge, D. Houston, J. Thorner, B. Errede, and H. G.
Dohlman. 2003. Regulation of Ste7 ubiquitination by Ste11
phosphorylation and the Skp1-Cullin-F-box complex. \emph{J. Biol.
Chem.} 278:22284-22289.\\
 60. Garrison, T. R., Y. Zhang, M. Pausch,
D. Apanovitch, R. Aebersold, and H. G. Dohlman. 1999. Feedback
phosphorylation of an RGS protein by MAP kinase in yeast. \emph{J.
Biol. Chem.} 274:36387-36391. \\

61. Goldbeter, Albert and Danniel E. Koshland. 1981. An amplified
sensitivity arising from covalent modification in biological
systems. \emph{Proc. Nati Acad. Sci. USA}. 78:6840-6844.\\

62. Wang, Y., and E. A. Elion. 2003. Nuclear export and plasma
membrane recruitment of the Ste5 scaffold are coordinated with
oligomerization and association with signal transduction components.
\emph{Mol. Biol. cell.} 14:2543-2558. \\
63. Yablonski, D., I. Marbach, and A. Levitzki. 1996. Dimerization
of Ste5, a mitogen-activated protein kinase cascade scaffold
protein, is required for signal transduction. \emph{Proc. Natl.
Acad. USA.} 93:13864-13869.
\end{flushleft}

\newpage

  \begin{table}[H]
\tabcolsep 1mm \caption{Coefficients in Hill function }
\begin{center}
\begin{tabular}{p{2.3cm}p{2.3cm}p{2.3cm}p{2.3cm}p{2.3cm}p{2.3cm}p{2.3cm}}
  \hline
  Coefficient  & Ste2a & $G$ activation  & Ste11pp & Ste7pp & Fus3pp \\
  n &0.9&1.0&1.2 &1.2 &1.2  \\
  th (nM)& 9.1&4.8&4.0 &4.8 &4.8  \\ \hline
\end{tabular}
\end{center}
Note: simulation results with the whole model.
\label{1}
\end{table}
\begin{table}[H]
\tabcolsep 1mm \caption{Hill coefficient n when negative feedback
are added}
\begin{center}
\begin{tabular}{p{7cm}p{2.5cm}p{2.5cm}p{2.5cm}}
 \hline
 added feedback & Ste11pp & Ste7pp & Fus3pp \\
 none &2.1&1.8&1.9 \\
 transcription of Sst2 & 1.8&1.5 &1.7 \\
 transcription of MAPK-P &2.0 &1.7&1.8 \\
 degradation of $Ste2_{act}$ &1.8&1.4&1.8 \\
 degradation of Ste11 &1.3&1.2&1.8 \\
 degradation of Ste7pp &2.1&1.7&1.7 \\
 hyper-phosphorylation of Ste7pp &1.8&1.6&1.4 \\
 all the six feedback &1.2&1.2&1.2 \\
 \hline
 \end{tabular}
 \end{center}
 \label{2}
 \end{table}
\begin{table}[H]
\tabcolsep 1mm \caption{Hill Coefficient n when concentration
changes}
\begin{center}
\begin{tabular}{p{7cm}p{2.5cm}p{2.5cm}p{2.5cm}}
  \hline
concentration changes&Ste11pp&Ste7pp&Fus3pp\\
$[Ste11]\times10$ &1.2&1.2&1.3\\
$[Ste7]\times10$ &1.3&1.6&1.2\\
$[Fus3]\times10$ &1.2&1.2&1.2\\
$[Ste11]\times10$, $[Ste5]\times10$ & 1.7 &1.8 & 1.4 \\
$[Ste7]\times10$, $[Ste5]\times10$ &1.8 &2.5 &2.8 \\
$[Fus3]\times10$, $[Ste5]\times10$ &1.4 &1.7 &1.7 \\
 $[MAPKKK-P]/10$ &1.7&1.7&1.4\\
 $[MAPKK-P]/10$&1.2&1.7&1.2\\
 $[MAPK-P]/10$&1.2&1.3&2.1\\\hline
\end{tabular}
\end{center}
\label{3}
\end{table}
\begin{table}[H]\tabcolsep 1mm
 \caption{Influential Parameters}
\begin{center}
\begin{tabular}{p{3cm}p{5cm}p{7cm}}
  \hline
  Parameter & $D=\sqrt{\overline{(\frac{\triangle [Fus3pp]}{[Fus3pp]})^{2}}}$ & Related reaction \\
  k1 &  0.519 & Ste2+$\alpha\underset{{k2}}{\overset{{k1}}{\rightleftharpoons}}Ste2_{act}$\\
  k2 &  0.266 &  \\
  k6 &  0.167 &  $\xrightarrow[{k6}]{{k4,k5,ste12a}}Ste2\xrightarrow{{k7}}$
  \\
  $**off_{K}$ & 0.156 & Fus3pp gets off from the scaffold protein \\
  $**off_{KK}'$  & 0.150 & Ste7pp gets off from the scaffold protein \\\hline
\end{tabular}
\end{center}
\label{4}
\end{table}

\newpage

\begin{flushleft}
\textbf{FIGURE LEGENDS}
\end{flushleft}

Figure 1. Spacial structure of the mating pathway.\\

Figure 2. A. 27 solution-located scaffold complexes. B. 27 membrane-located scaffold complexes. The diamond on the upleft corner indicates the G$\beta\gamma$Ste20 complex\\

Figure 3. Scaffold-dependent reactions. \\

Figure 4. Time course of the  G protein cycle activation. A. G
protein activation. The values are normalized to the maximum
concentration around 30 s. Result from simulation is shown
  in solid line and experiment data (12) are plotted in circles with error-bars. B. G protein
  activation in\emph{ $Ste2^{300\triangle}$} cells, the wild types and cycloheximide treated cells (experiment
  data are from (12)) .
  C. The time-course of Binding Ste20 to G$\beta\gamma$. The values are normalized
  to the maximum concentration (experiment data are from (24)). D. The recruitment of scaffold protein
  Ste5 (circles) which is the sum of scaffolds in the solution and scaffolds at the membrane, the activation of MAPK cascade components Ste11pp (MAPKKKK)(dotted line),
  Ste7pp (MAPKK) (dashed line) and Fus3pp (MAPK)(solid line). E. Downstream responses
  to $\alpha-factor$ induction: activated Ste12 (solid line), $Far1pp\cdot Cdc28$ (dashed line),
  and $Far1pp\cdot G\beta\gamma$.\\

Figure 5. A. Predicted dose-response curves of recruitment of
  scaffold protein Ste5 (dashed line) and activation of Fus3 (MAPK)
  (solid line). B. The dependence of Fus3pp on the concentration of
  scaffold proteins Ste5, with different concentration of $\alpha$-factor.
  C. Predicted dose-response curves for mutants in which MAPK-P is
  under-expressed. All the values are normalized. In all above simulation,
  Cells are treated with indicated with $\alpha-factor$ for 20 min. \\

Figure 6. Dephosphorylation prevents improper signal to leak in.
  A. All Ste11 is dual phosphorylated at $t=0$, and MAPKKK-P is shut off. This
  simulates the condition in which signal in invasive growth pathway
  is on. B. All Ste11 is dual phosphorylated at $t=0$, and MAPKK-P is shut off. The
  activation of Fus3 is still repressed down except for the small
  pulse at the very beginning (Inset graph). Dashed line indicates
  the activation of Fus3 when MAPK-P is eliminated.\\

Figure 7. Desensitization. A. Time-course of activation of Fus3
(MAPK) in a wild type
  cell (solid line) and in a mutant (dashed line). B. Effect of feedback hyper-phosphorylation of Ste7 in the
  scaffold. Predicted time-course activation of Fus3 for wild type
  (solid line) and for mutants (dashed line and dotted line).\\

Figure 8. Dose-response curves for key components in mating pathway
 in TMY101 cells: activated Ste2 (dotted line for simulation,
 down-triangles for experiment data from (12)); G protein activation, i.e.
 the sum of  G$\alpha$-GTP and G$\alpha$-GDP (dashed line for simulation, circles for
 experiment data from (12)); Fus3pp (MAPK) (solid line for simulation and
 crosses for experiment data from (45)).\\

Figure 9. Response curve of Ste12 to $\alpha$-factor for: A.
 different Sst2 expression level; B. different G$\beta\gamma$ expression level.
 To simulate the $2\times G\beta\gamma$ cells, we separately add another
 1000nM G$\beta\gamma$ at t=0 min to double the total concentration of
 G$\beta\gamma$.  Cells are treated with $\alpha$-factor for 60 min.
 Note that the experiment data are shifted left by more than 100-fold because \emph{Bar1}
 is deleted in TMY101 cells. Experiment data are from (8).\\

\newpage

\begin{figure}[H]
  \includegraphics[width=6in]{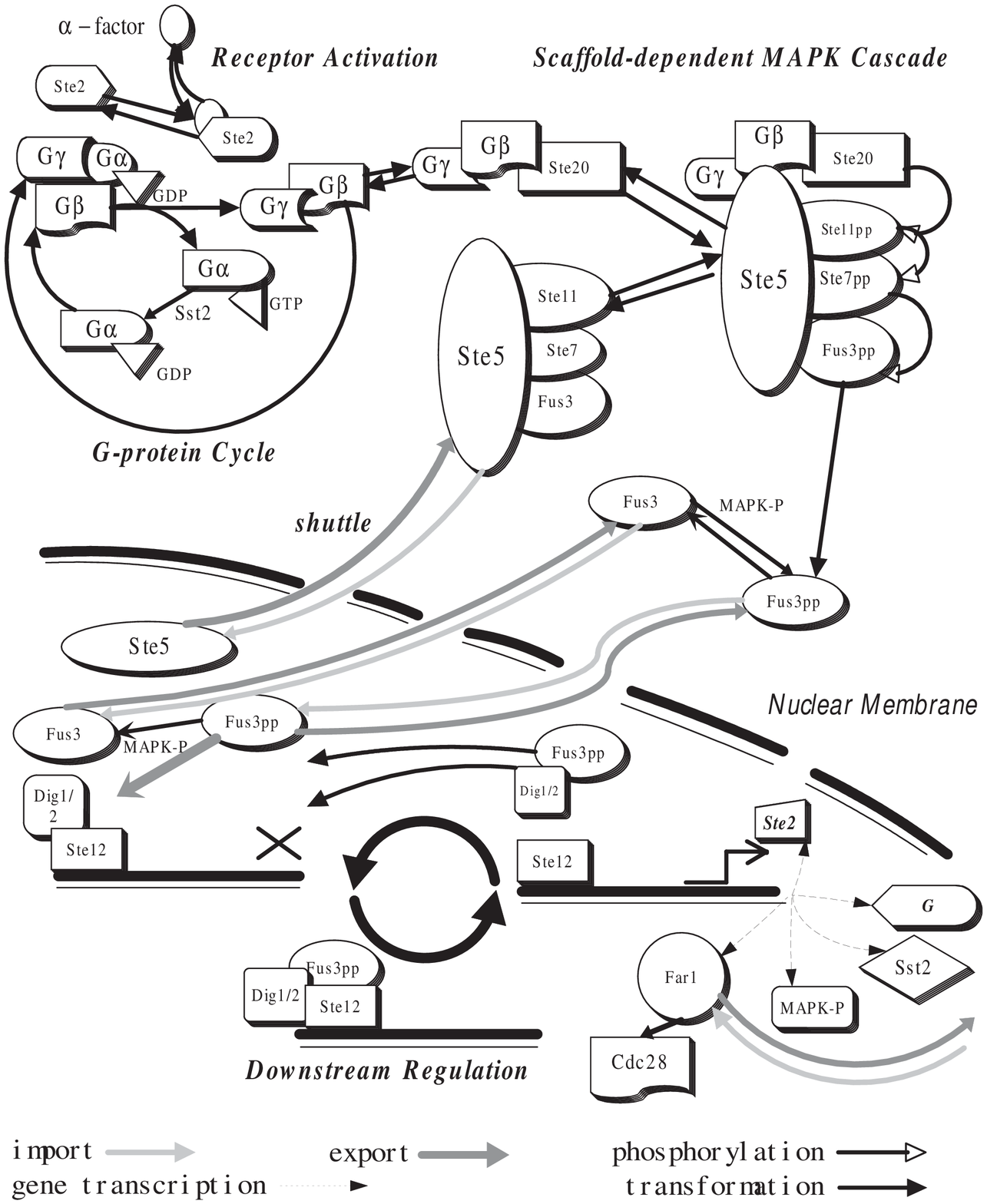}
  \caption{}
  \label{fig1}
\end{figure}

\newpage
\begin{figure}[H]

  \includegraphics[width=6in]{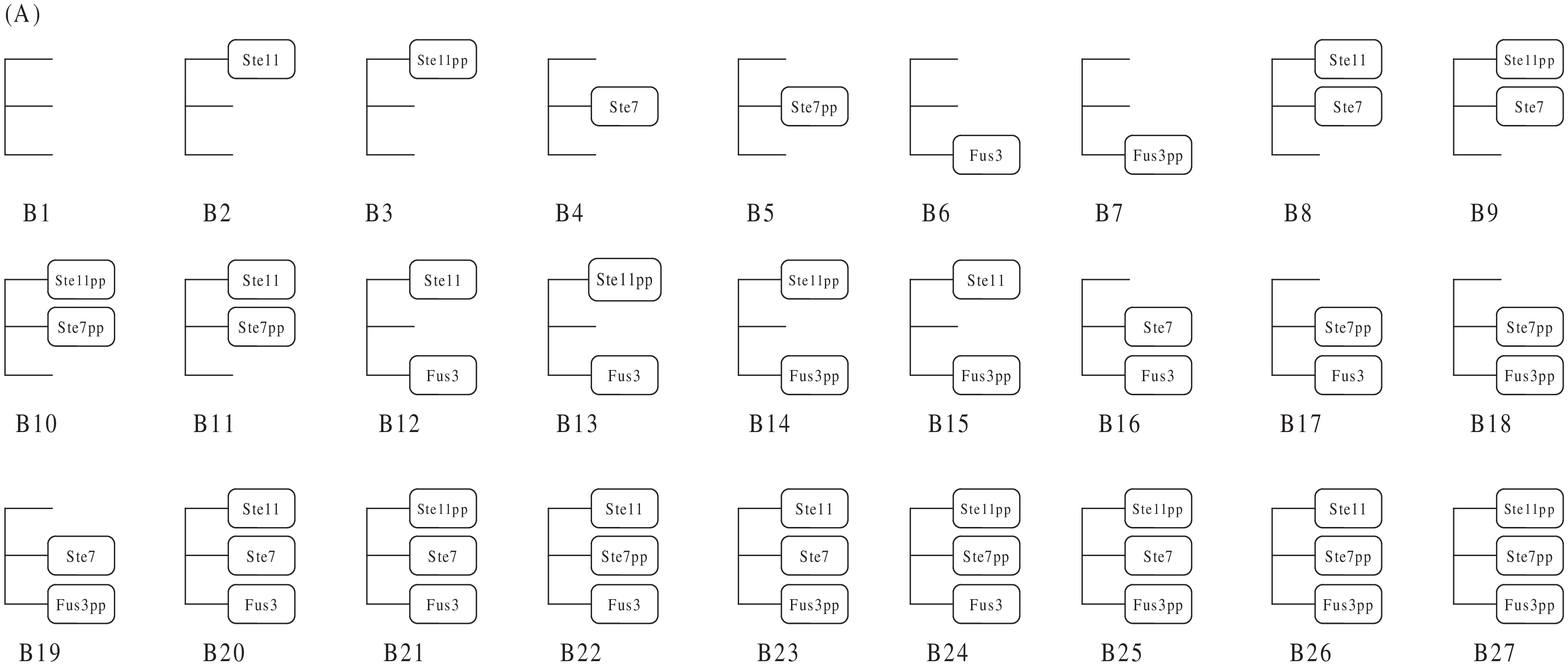}\\
  \includegraphics[width=6in]{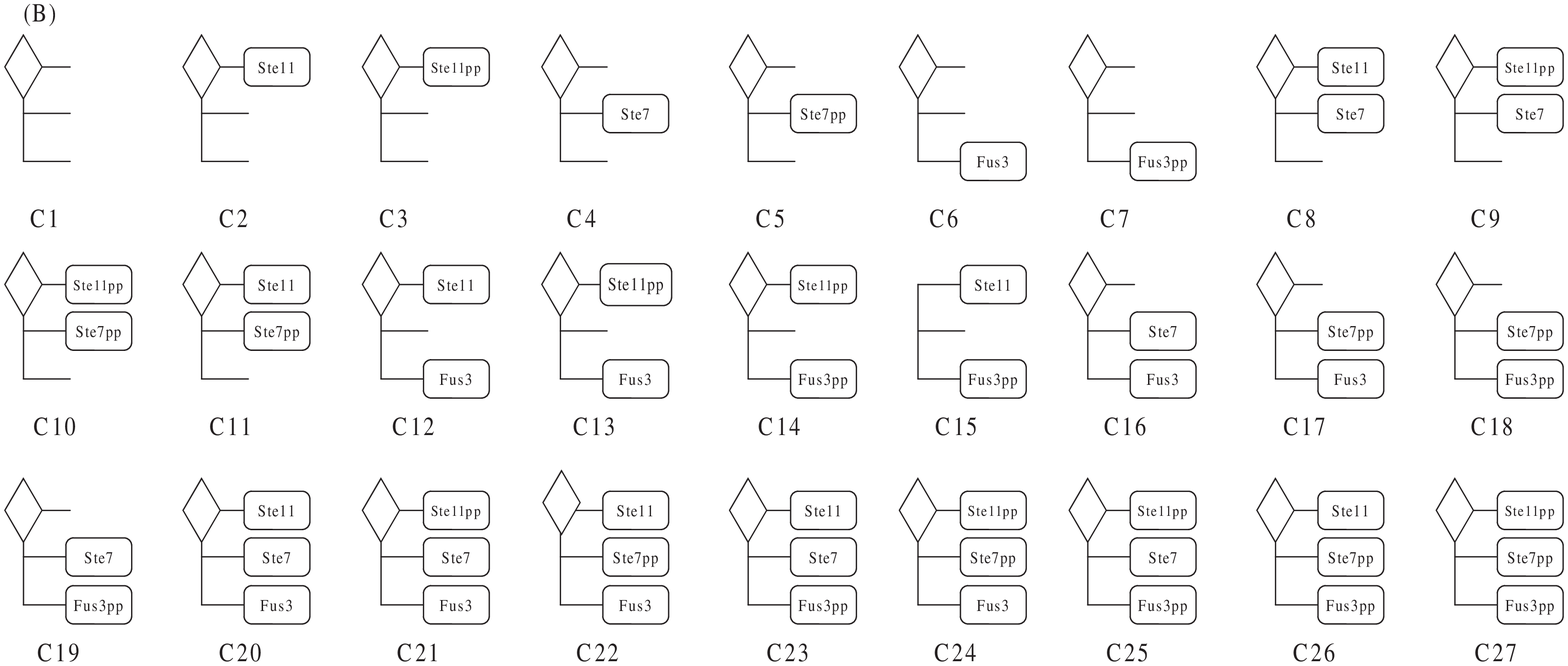}
  \caption{}

  \label{fig2}
\end{figure}

\newpage
\begin{figure}[H]
\begin{center}
  \includegraphics[width=4in]{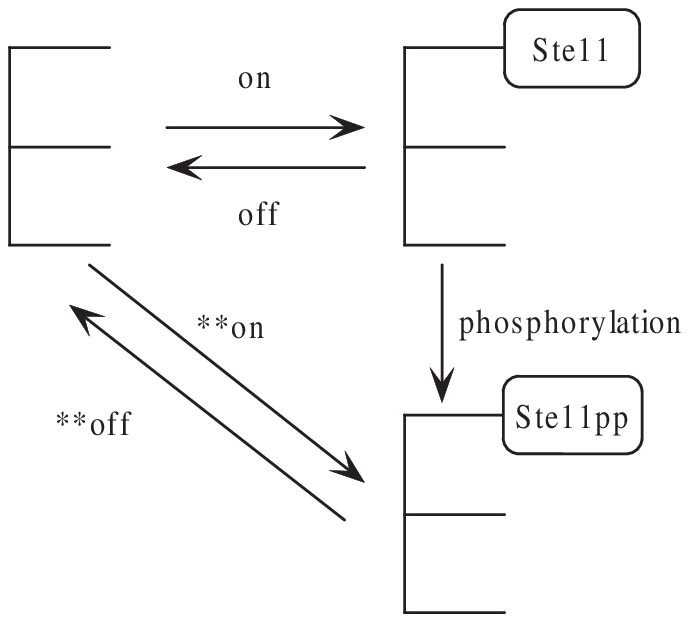}
  \caption{}
  \end{center}
  \label{fig3}
\end{figure}

\newpage
\begin{figure}[H]
 \center{
  \includegraphics[width=1.9in]{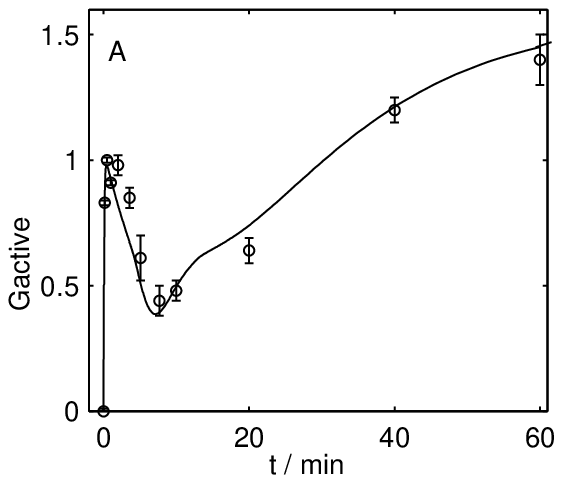}\quad
  \includegraphics[width=1.9in]{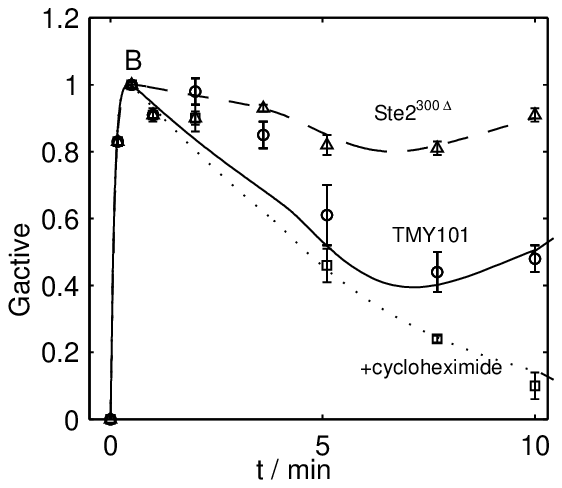}}\\
\mbox{
 \includegraphics[width=1.9in]{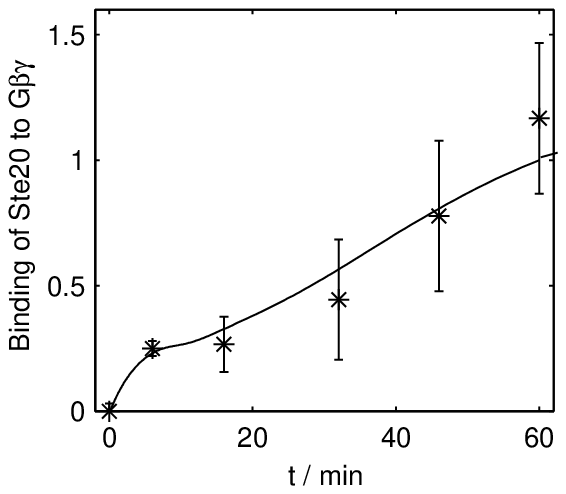}
 \includegraphics[width=2.2in]{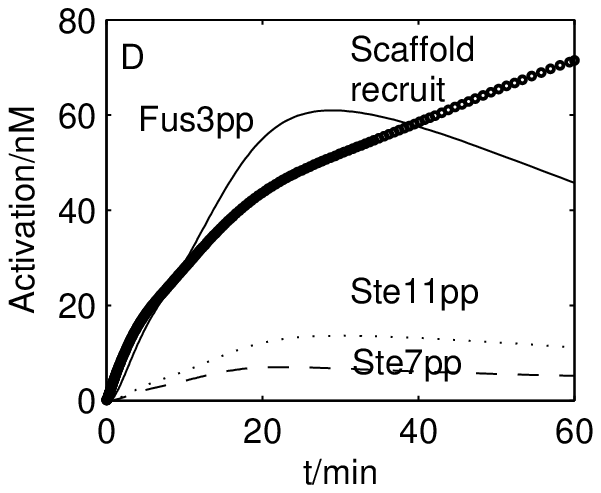}
 \includegraphics[width=2.2in]{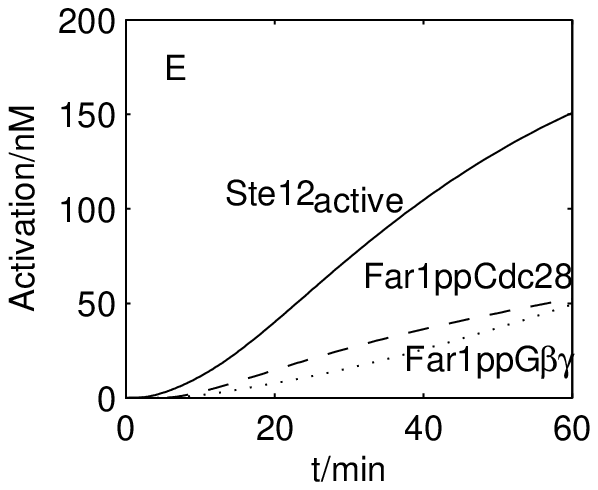}}
  \caption{}
 \label{fig4}
\end{figure}

\newpage
\begin{figure}[H]
  \includegraphics[width=2in]{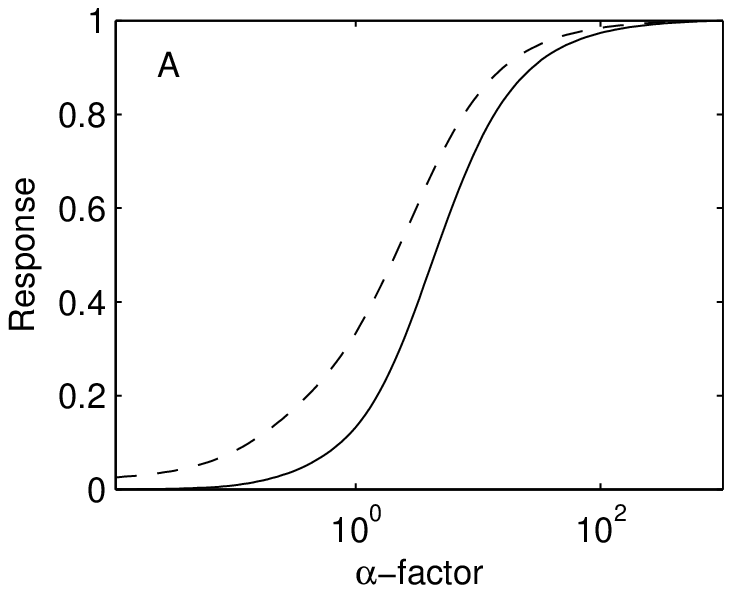}
\includegraphics[width=2in]{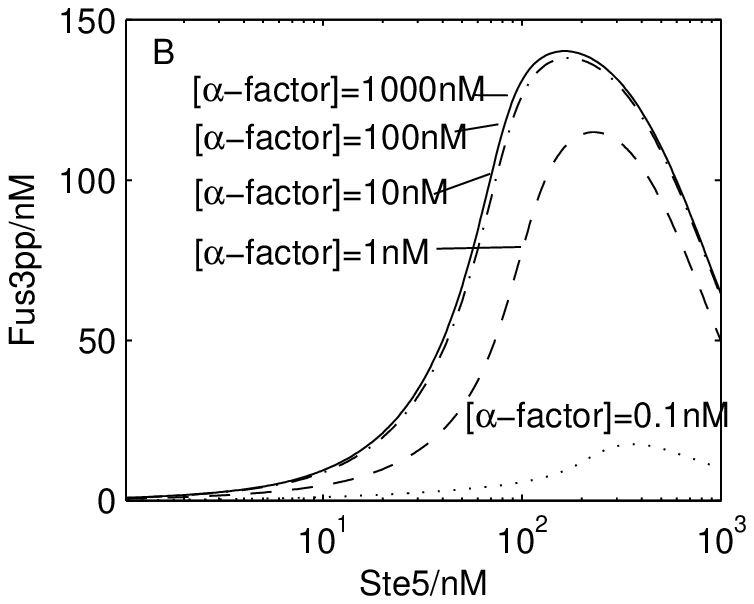}
  \includegraphics[width=2in]{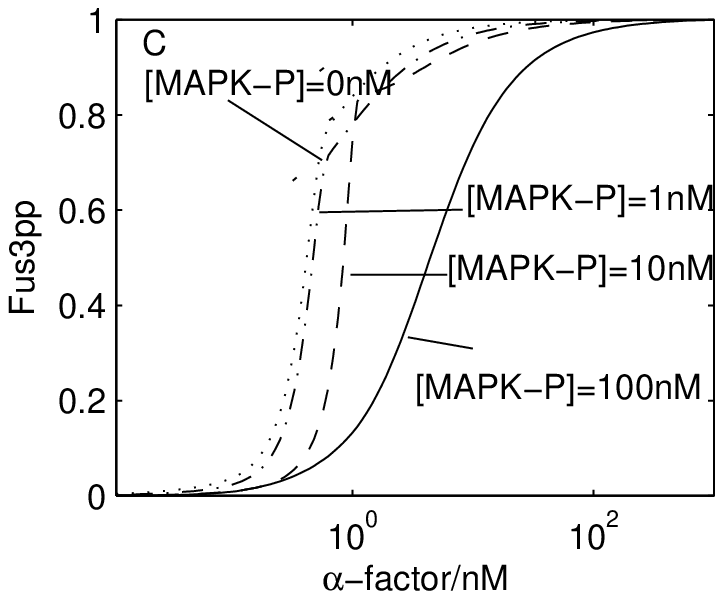}
  \caption{ }
  \label{fig5}
\end{figure}

\newpage
\begin{figure}[H]
 \mbox{\includegraphics[width=3in]{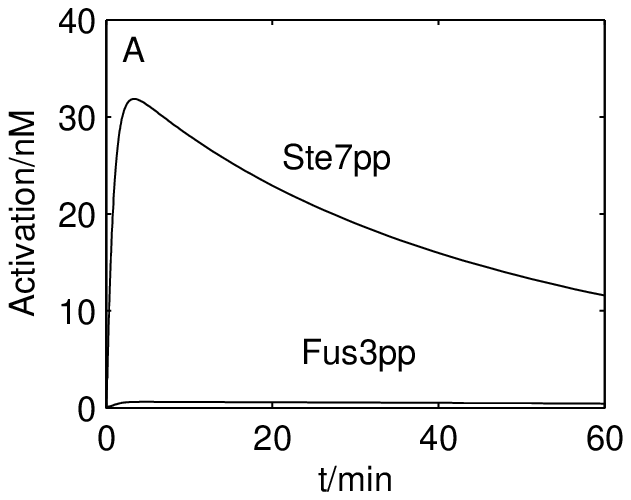}
  \includegraphics[width=3in]{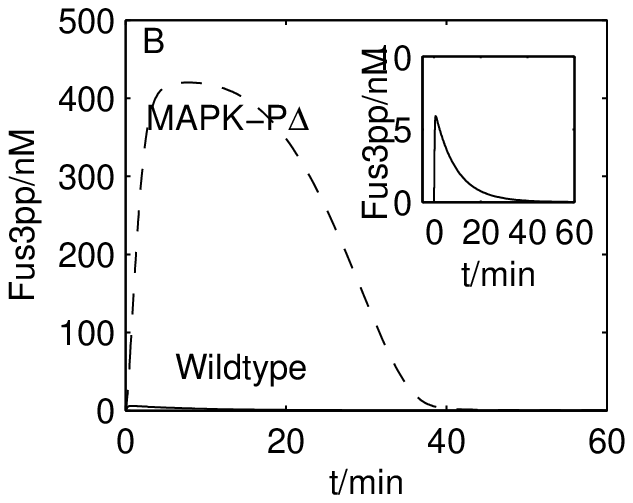}}
  \label{fig6}
  \caption{}
\end{figure}

\newpage
\begin{figure}[H]
  \includegraphics[width=3in]{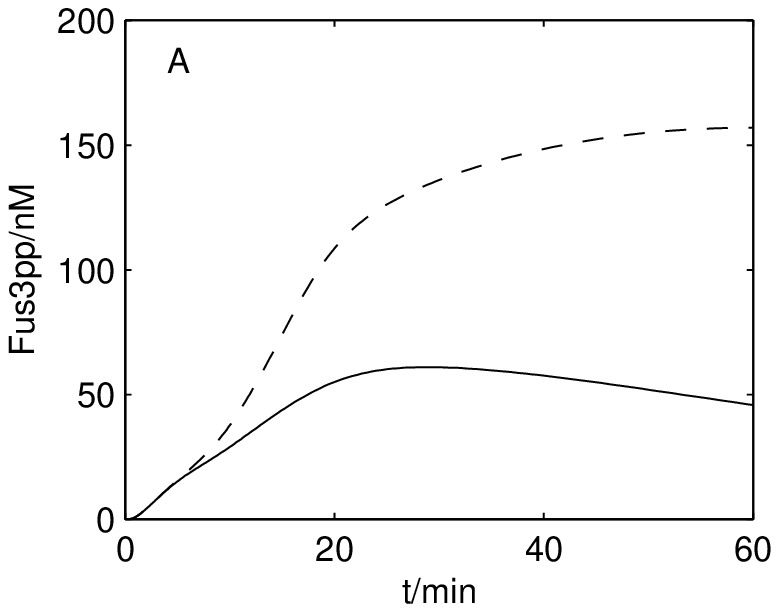}
  \includegraphics[width=3in]{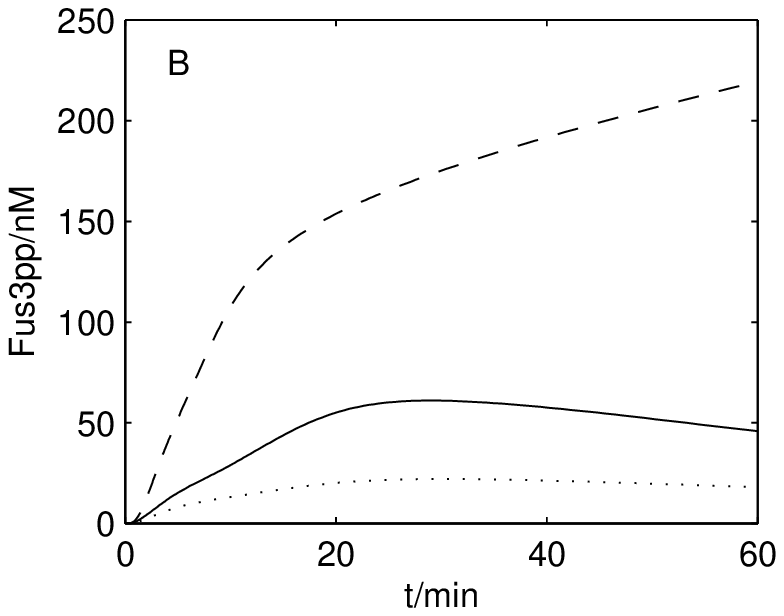}
  \caption {}
  \label{fig7}
\end{figure}

\newpage
\begin{figure}[H]
 \begin{center}
 \includegraphics[width=4in]{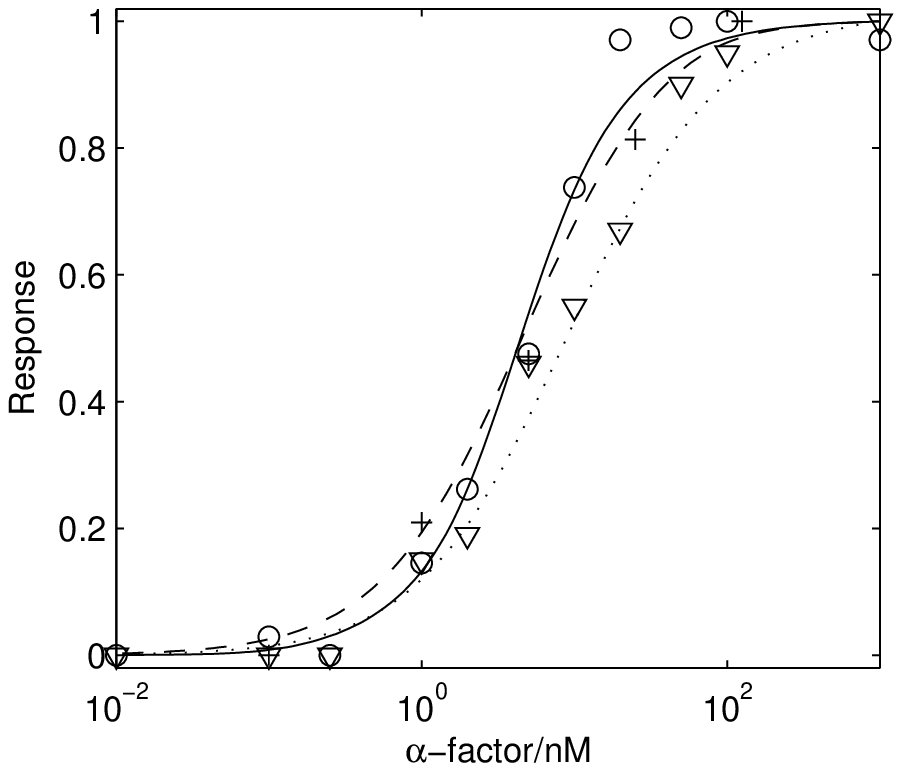}
 \end{center}
 \caption{}
  \label{fig8}
\end{figure}

\newpage
\begin{figure}[H]
\mbox{
\includegraphics[width=3in]{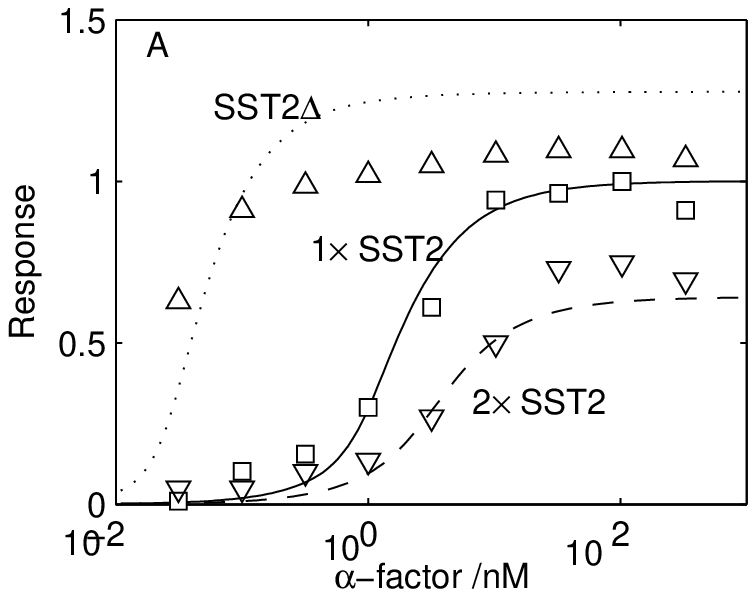}
 \includegraphics[width=3in]{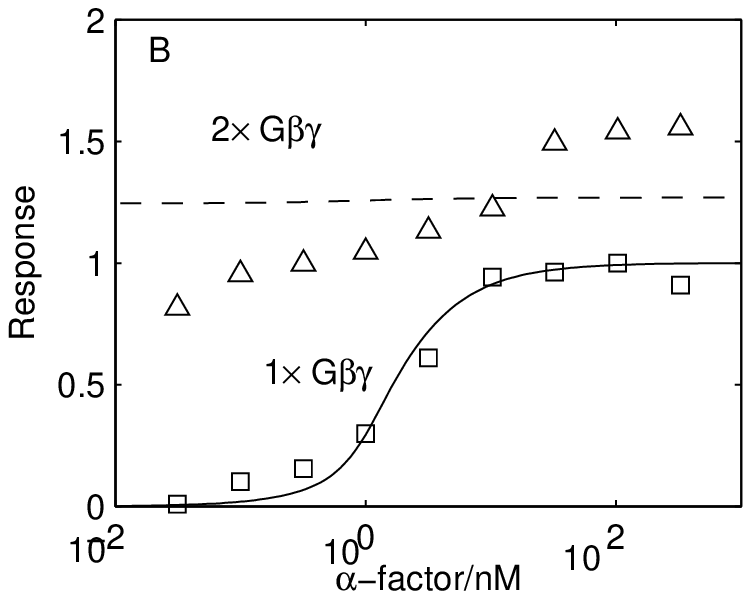}}
 \caption{ }
 \label{fig9}
\end{figure}

\end{document}


\title{SUPPLEMENTAL MATERIAL}
\maketitle
\begin{table}[!h]
\caption{parameters in G-protein cycle}
\begin{center}
\begin{tabular}{|c|c|p{9cm}|}
  \hline
  parameter & value & References and Notes \\ \hline
  k1 & $0.12 nM^{-1}min^{-1}$ & $2\times 10^6 M^{-1}s^{-1}$ (12)\\\hline
  k2 & $0.6 min^{-1}$& 0.01 $s^{-1}$ (12)\\\hline
  k3 & $0.24 min^{-1}$& $4\times 10^{-3} s^{-1}$ (12)\\\hline
  k4 & $320 nM\cdot min^{-1}$ & fit to the time-course of G protein activation\\\cline{1-2}
  k5 & 7 nM &   (12).(See Fig.2A ) \\\hline
  k6 & $30 nM\cdot min^{-1}$& $4 (molecules/cell)s^{-1}$ (12) \\\hline
  k7 & $0.024 min^{-1} $& $4\times 10^{-4} s^{-1}$  (12)\\\hline
  k8 & $0.0048 min^{-1}$& $10^{-5} (molecules/cell)^{-1}s^{-1}$ (12)\\\hline
  k9 & $115 nM\cdot min^{-1}$ & fit to the time-course of G protein activation \\\cline{1-2}
  k10 & 60 nM & (12)(See Fig.2A), \\\cline{1-2}
  k11 & $20 nM\cdot min^{-1}$ & and experimental data that G protein increased from \\\cline{1-2}
  k12 & $ 0.08 min^{-1}$& 8000 molecules/cell to 12000 molecules/cell (8)\\\hline
  k13 & $0.24 min^{-1}$& $4\times 10^{-3} s^{-1}$ (12) \\\hline
  k14 &$ 0.025 nM^{-1}min^{-1}$& $0.11 s^{-1}$ (12)\\\hline
  k15 & $480nM^{-1}min^{-1}$ & $1 (molecule/cell)^{-1}s^{-1} $(12)\\
  \hline
\end{tabular}
\end{center}
\end{table}

\begin{table}[!h]
\caption{parameters in recruitment of scaffolds }
\begin{center}
\begin{tabular}{|c|c|p{7cm}|}
  \hline
  parameter & value & References and Notes \\ \hline
  k16 & $0.05 nM^{-1}min^{-1}$ & $t_{1/2}=8.22s$ for Ste5s' recovery  \\\cline{1-2}
  k17 & $5 min^{-1}$ & at shmoo tips (49)\\\hline
  k18 & $0.00007 nM^{-1}min^{-1}$ & fit to (24)\\\cline{1-2}
  k19 & $0.001 min^{-1}$& (See Fig.2C. )  \\\hline
  k23 & $30 min^{-1}$& $t_{1/2}=2 s$ for the recovery of ste5 in the nucleus. (49)  \\
  \hline
\end{tabular}
\end{center}
\end{table}

\begin{table}[!h]
\tabcolsep 5mm
\caption{dephosphorylation}
\begin{center}
\begin{tabular}{|c| p{2cm}|p{1cm} |p{1cm}|p{2cm}|p{1cm}|p{1cm}|}  \hline
protein& a1 ($ nM^{-1}min^{-1}$)&d1 ($min^{-1}$)&p1 ($min^{-1}$)&a2 ($ nM^{-1}min^{-1}$)&d2 ($min^{-1}$)&p2 ($min^{-1}$) \\\hline
Ste11(MAPKKK)& 0.5 & 24 & 10&  1& 24& 10 \\ \hline
Ste7(MAPKK) & 0.5 & 24 & 10& 0.5& 24 & 10\\ \hline
Fus3(MAPK)& 0.2& 10& 30 & 0.4 & 20 & 30   \\ \hline
\end{tabular}
\end{center}
\end{table}

\begin{table}[!h]
\tabcolsep 5mm
\caption{phosphorylation in cytosol}
\begin{center}
\begin{tabular}{|c|p{2cm}|p{1cm}|p{1cm}|p{2cm}|p{1cm}|p{1cm}| }
  \hline
protein& a3 ($ nM^{-1}min^{-1}$)&d3 ($min^{-1}$)&p3 ($min^{-1}$)&a4 ($ nM^{-1}min^{-1}$)&d4 ($min^{-1}$)&p4 ($min^{-1}$) \\\hline
Ste7(MAPKK)& 1 & 36&10&  1  & 36 &10 \\ \hline
Fus3(MAPK) & 0.1 & 36 & 10 &0.1& 36& 10\\ \hline
  \end{tabular}
  \end{center}
Notes: Equations of reaction in this part are all in the following
form:
$S+E\underset{d_{n}}{\overset{a_{n}}{\rightleftharpoons}}SE\xrightarrow{p_{n}}S^{*}+E$.
\begin{flushleft}
As for the underset n: \\
1: dephosphorylation of the once
phosphorylated kinase; 2: dephosphorylation of the twice
phosphorylated kinase; 3: once phosphorylation; 4: twice
phosphorylation. Parameters are estimated in analogy to Ref.11.
\end{flushleft}
\end{table}
\begin{table}[!h]
\tabcolsep 5mm
\caption{Other reactions in the cytosol}
\begin{center}
\begin{tabular}{|c|c|l|}
  \hline
parameter&value&References \\ \hline
   k24& 1.2 $ nM^{-1}min^{-1}$& Ref.23 \\ \hline
  k25 & 24$ min^{-1}$ & Ref.23 \\ \hline
  k26 &0.0015 $ nM^{-1}min^{-1}$ &  Ref.40  \\ \hline
  k27& 0.1 $min^{-1}$& Ref.41\\ \hline
\end{tabular}
\end{center}
\end{table}

\begin{figure}
\includegraphics[width=11cm]{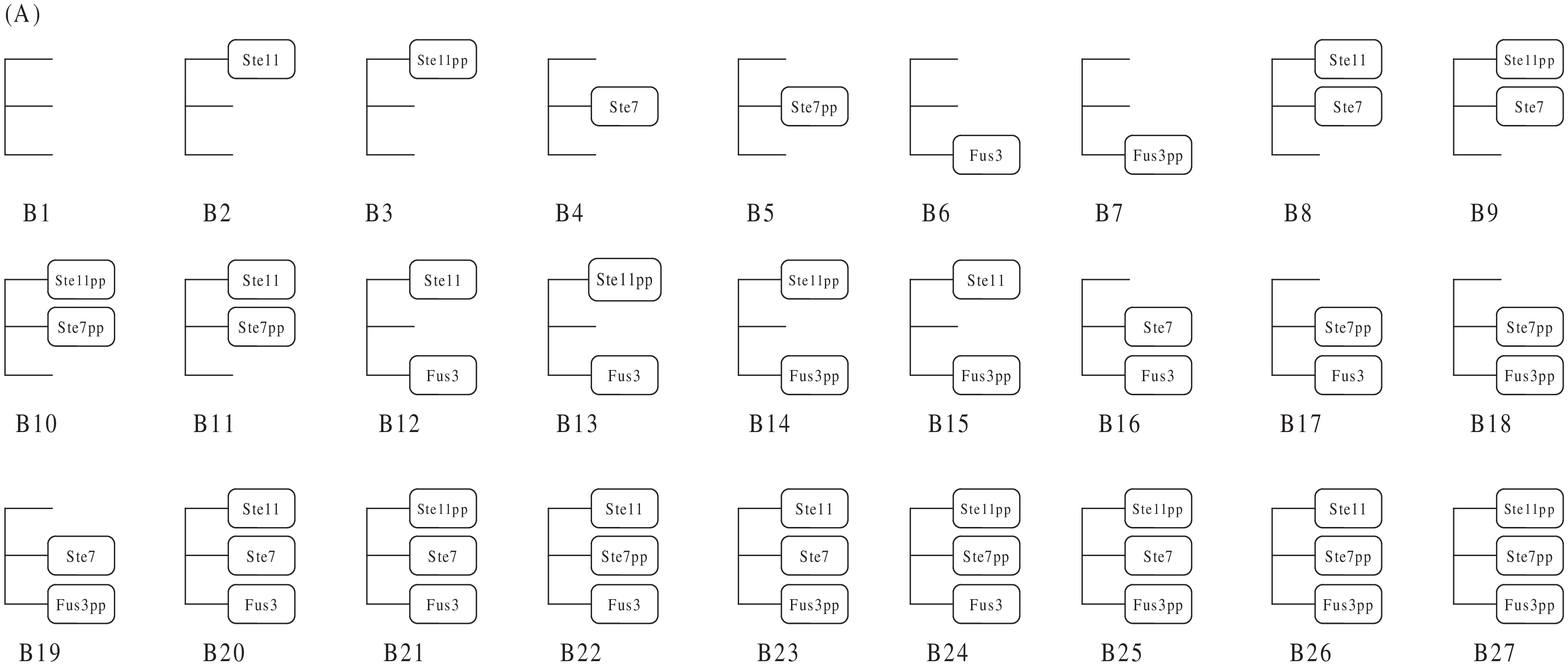}\\
  \includegraphics[width=11cm]{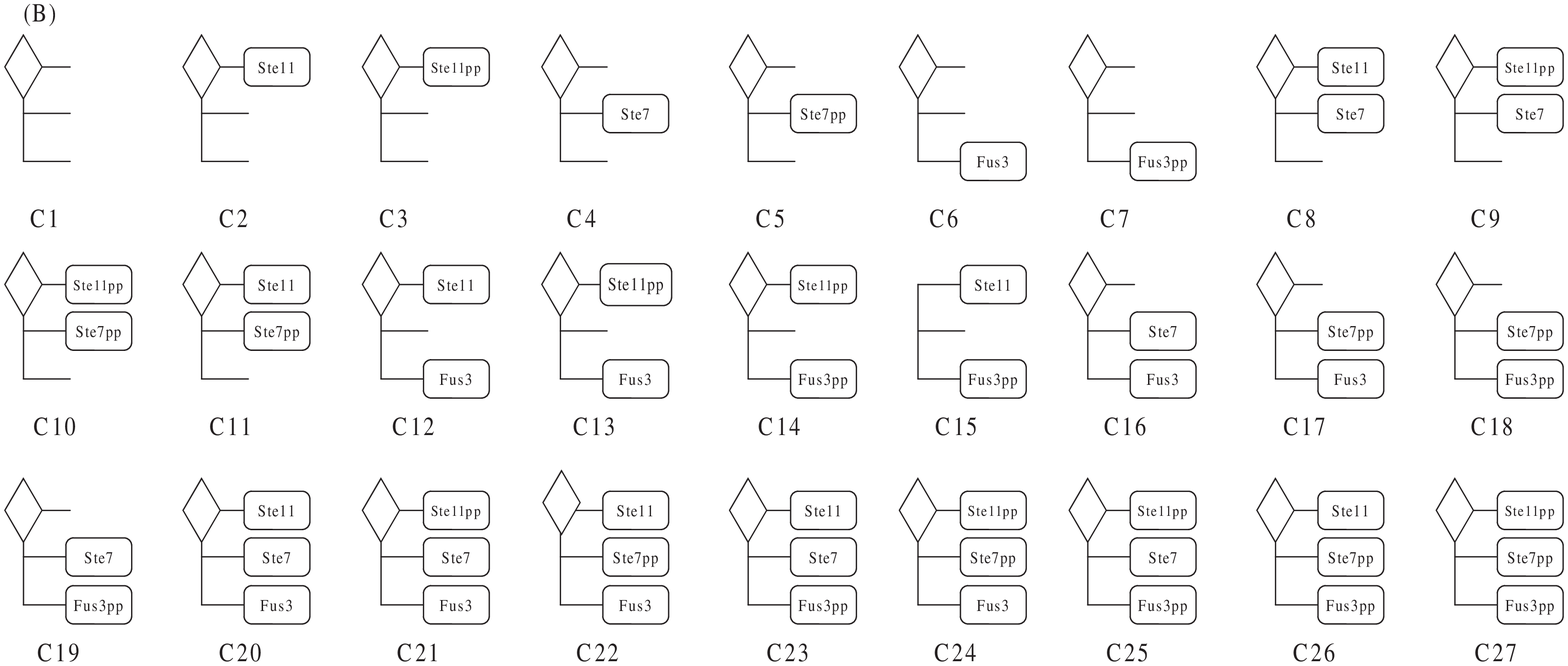}\\
  \caption{\label{Fig.}27 kinds of scaffold-kinase complexes in solution and 27 kinds of scaffold-kinase complexes at the membrane}
\end{figure}

\footnotesize
\[\begin{array}{lll}
B1 + Ste11\underset{{}}{\overset{{}}{\rightleftharpoons}}B2 &
 B1 + Ste11pp\underset{{}}{\overset{{}}{\rightleftharpoons}}B3  &
  B1 + Ste7\underset{{}}{\overset{{}}{\rightleftharpoons}}B4 \\
  B1 + Ste7pp\xleftarrow{{}}B5&
  B1 + Fus3\underset{{}}{\overset{{}}{\rightleftharpoons}}B6 &
  B1 + Fus3pp\xleftarrow{{}}B7 \\
B2 + Ste7\underset{{}}{\overset{{}}{\rightleftharpoons}}B8 &
  B2 + Ste7pp\xleftarrow{{}}B11  &
  B2 + Fus3\underset{{}}{\overset{{}}{\rightleftharpoons}}B12 \\
  B2 + Fus3pp\xleftarrow{{}}B15&
  B3 + Ste7\underset{{}}{\overset{{}}{\rightleftharpoons}}B9  &
  B3 + Ste7pp\xleftarrow{{}}B10\\
  B3 + Fus3\underset{{}}{\overset{{}}{\rightleftharpoons}}B13  &
  B3 + Fus3pp\xleftarrow{{}}B14 &
  B4 + Ste11\underset{{}}{\overset{{}}{\rightleftharpoons}}B8  \\
  B4 + Ste11pp\underset{{}}{\overset{{}}{\rightleftharpoons}}B9 &
  B4 + Fus3\underset{{}}{\overset{{}}{\rightleftharpoons}}B16  &
  B4 + Fus3pp\xleftarrow{{}}B19 \\
  B5 + Fus3pp\underset{{}}{\overset{{}}{\rightleftharpoons}}B10  &
  B5 + Ste11\underset{{}}{\overset{{}}{\rightleftharpoons}}C11
  \hfill &
  B5 + Fus3\underset{{}}{\overset{{}}{\rightleftharpoons}}B17 \\
  B5 + Fus3pp\xleftarrow{{}}B18 &
  B6 + Ste11\underset{{}}{\overset{{}}{\rightleftharpoons}}B12 &
  B6 + Ste11pp\underset{{}}{\overset{{}}{\rightleftharpoons}}B13 \\
  B6 + Ste7\underset{{}}{\overset{{}}{\rightleftharpoons}}B16 &
  B6 + Ste7pp\xleftarrow{{}}B17 &
  B7 + Ste11pp\underset{{}}{\overset{{}}{\rightleftharpoons}}B14 \\
  B7 + Ste11\underset{{}}{\overset{{}}{\rightleftharpoons}}B15 &
  B7 + STe7pp\xleftarrow{{}}B18 &
  B7 + Ste7\underset{{}}{\overset{{}}{\rightleftharpoons}}B19 \\
B8 + Fus3\underset{{}}{\overset{{}}{\rightleftharpoons}}B20 &
  B8 + Fus3pp\xleftarrow{{}}B23 &
  B9\xrightarrow{{}}B10 \\
  B9 + Fus3\underset{{}}{\overset{{}}{\rightleftharpoons}}B21 &
  B9 + Fus3pp\xleftarrow{{}}B25 &
B10 + Fus3\underset{{}}{\overset{{}}{\rightleftharpoons}}B24 \\
  B10 + Fus3pp\xleftarrow{{}}B27 &
  B11 + Fus3\underset{{}}{\overset{{}}{\rightleftharpoons}}B22 &
  B11 + Fus3pp\xleftarrow{{}}B26 \\
B12 + Ste7\underset{{}}{\overset{{}}{\rightleftharpoons}}B20&
  B12 + Ste7pp\xleftarrow{{}}B22&
  B13 + Ste7\underset{{}}{\overset{{}}{\rightleftharpoons}}B21 \\
  B13 + Ste7pp\xleftarrow{{}}B24 &
B14 + Ste7\underset{{}}{\overset{{}}{\rightleftharpoons}}B25 &
  B14 + Ste7pp\xleftarrow{{}}B27 \\
  B15 + Ste7\underset{{}}{\overset{{}}{\rightleftharpoons}}B23 &
  B15 + Ste7pp\xleftarrow{{}}B26 &
  B16 + Ste11pp\underset{{}}{\overset{{}}{\rightleftharpoons}}B21 \\
  B16 + Ste11\underset{{}}{\overset{{}}{\rightleftharpoons}}B20 &
  B17\xrightarrow{{}}B18 &
  B17 + Ste11pp\underset{{}}{\overset{{}}{\rightleftharpoons}}B24 \\
  B17 + Ste11\underset{{}}{\overset{{}}{\rightleftharpoons}}B22&
  B18 + Ste11pp\underset{{}}{\overset{{}}{\rightleftharpoons}}B27 &
  B18 + Ste11\underset{{}}{\overset{{}}{\rightleftharpoons}}B26 \\
  B19 + Ste11pp\underset{{}}{\overset{{}}{\rightleftharpoons}}B25 &
  B19 + Ste11\underset{{}}{\overset{{}}{\rightleftharpoons}}B23 &
B21\xrightarrow{{}}B24 \\
B22\xrightarrow[{}]{{}}B26 & B24\xrightarrow{{}}B27 &
  B25\xrightarrow[{}]{{}}B27 \\
C1 + Ste11\underset{{}}{\overset{{}}{\rightleftharpoons}}C2 &
 C1 + Ste11pp\underset{{}}{\overset{{}}{\rightleftharpoons}}C3  &
  C1 + Ste7\underset{{}}{\overset{{}}{\rightleftharpoons}}C4 \\
  C1 + Ste7pp\xleftarrow{{}}C5&
  C1 + Fus3\underset{{}}{\overset{{}}{\rightleftharpoons}}C6 &
  C1 + Fus3pp\xleftarrow{{}}C7 \\
  C2\xrightarrow{{}}C3 &
  C2 + Ste7\underset{{}}{\overset{{}}{\rightleftharpoons}}C8 &
  C2 + Ste7pp\xleftarrow{{}}C11  \\
  C2 + Fus3\underset{{}}{\overset{{}}{\rightleftharpoons}}C12 &
  C2 + Fus3pp\xleftarrow{{}}C15&
  C3 + Ste7\underset{{}}{\overset{{}}{\rightleftharpoons}}C9  \\
  C3 + Ste7pp\xleftarrow{{}}C10&
  C3 + Fus3\underset{{}}{\overset{{}}{\rightleftharpoons}}C13  &
  C3 + Fus3pp\xleftarrow{{}}C14 \\
  C4 + Ste11\underset{{}}{\overset{{}}{\rightleftharpoons}}C8  &
  C4 + Ste11pp\underset{{}}{\overset{{}}{\rightleftharpoons}}C9 &
  C4 + Fus3\underset{{}}{\overset{{}}{\rightleftharpoons}}C16  \\
  C4 + Fus3pp\xleftarrow{{}}C19 &
  C5 + Fus3pp\underset{{}}{\overset{{}}{\rightleftharpoons}}C10  &
  C5 + Ste11\underset{{}}{\overset{{}}{\rightleftharpoons}}C11 \hfill \\
  C5 + Fus3\underset{{}}{\overset{{}}{\rightleftharpoons}}C17 &
  C5 + Fus3pp\xleftarrow{{}}C18 &
  C6 + Ste11\underset{{}}{\overset{{}}{\rightleftharpoons}}C12 \\
  C6 + Ste11pp\underset{{}}{\overset{{}}{\rightleftharpoons}}C13 &
  C6 + Ste7\underset{{}}{\overset{{}}{\rightleftharpoons}}C16 &
  C6 + Ste7pp\xleftarrow{{}}C17 \\
  C7 + Ste11pp\underset{{}}{\overset{{}}{\rightleftharpoons}}C14 &
  C7 + Ste11\underset{{}}{\overset{{}}{\rightleftharpoons}}C15 &
  C7 + STe7pp\xleftarrow{{}}C18 \\
  C7 + Ste7\underset{{}}{\overset{{}}{\rightleftharpoons}}C19 &
  C8\xrightarrow{{}}C9 &
  C8 + Fus3\underset{{}}{\overset{{}}{\rightleftharpoons}}C20 \\
  C8 + Fus3pp\xleftarrow{{}}C23 &
  C9\xrightarrow{{}}C10 &
  C9 + Fus3\underset{{}}{\overset{{}}{\rightleftharpoons}}C21 \\
  C9 + Fus3pp\xleftarrow{{}}C25 &
C10\xleftarrow[{}]{{}}C11 &
  C10 + Fus3\underset{{}}{\overset{{}}{\rightleftharpoons}}C24 \\
  C10 + Fus3pp\xleftarrow{{}}C27 &
  C11 + Fus3\underset{{}}{\overset{{}}{\rightleftharpoons}}C22 &
  C11 + Fus3pp\xleftarrow{{}}C26 \\
  C12\xleftarrow[{}]{{}}C13 &
  C12 + Ste7\underset{{}}{\overset{{}}{\rightleftharpoons}}C20&
  C12 + Ste7pp\xleftarrow{{}}C22\\
  C13 + Ste7\underset{{}}{\overset{{}}{\rightleftharpoons}}C21 &
  C13 + Ste7pp\xleftarrow{{}}C24 &
  C14\xleftarrow[{}]{{}}C15 \\
  C14 + Ste7\underset{{}}{\overset{{}}{\rightleftharpoons}}C25 &
  C14 + Ste7pp\xleftarrow{{}}C27 &
  C15 + Ste7\underset{{}}{\overset{{}}{\rightleftharpoons}}C23 \\
  C15 + Ste7pp\xleftarrow{{}}C26 &
  C16 + Ste11pp\underset{{}}{\overset{{}}{\rightleftharpoons}}C21 &
  C16 + Ste11\underset{{}}{\overset{{}}{\rightleftharpoons}}C20 \\
  C17\xrightarrow{{}}C18 &
  C17 + Ste11pp\underset{{}}{\overset{{}}{\rightleftharpoons}}C24 &
  C17 + Ste11\underset{{}}{\overset{{}}{\rightleftharpoons}}C22\\
  C18 + Ste11pp\underset{{}}{\overset{{}}{\rightleftharpoons}}C27 &
  C18 + Ste11\underset{{}}{\overset{{}}{\rightleftharpoons}}C26 &
  C19 + Ste11pp\underset{{}}{\overset{{}}{\rightleftharpoons}}C25 \\
  C19 + Ste11\underset{{}}{\overset{{}}{\rightleftharpoons}}C23 &
  C20\xrightarrow{{}}C21 &
  C21\xrightarrow{{}}C24 \\
  C22\xrightarrow[{}]{{}}C24 &
  C22\xrightarrow[{}]{{}}C26 &
  C23\xrightarrow[{}]{{}}C25 \\
  C24\xrightarrow{{}}C27 &
  C25\xrightarrow[{}]{{}}C27 &
  C26\xrightarrow[{}]{{}}C27 \\
  B1 \underset{{}}{\overset{{}}{\rightleftharpoons}}C1 &
   B2 \underset{{}}{\overset{{}}{\rightleftharpoons}}C2 &
    B3 \underset{{}}{\overset{{}}{\rightleftharpoons}}C3 \\
     B4\underset{{}}{\overset{{}}{\rightleftharpoons}}C4 &
      B5 \underset{{}}{\overset{{}}{\rightleftharpoons}}C5 &
       B6 \underset{{}}{\overset{{}}{\rightleftharpoons}}C6 \\
B7 \underset{{}}{\overset{{}}{\rightleftharpoons}}C7 &
   B8 \underset{{}}{\overset{{}}{\rightleftharpoons}}C8 &
    B9 \underset{{}}{\overset{{}}{\rightleftharpoons}}C9 \\
     B10 \underset{{}}{\overset{{}}{\rightleftharpoons}}C10 &
     B11 \underset{{}}{\overset{{}}{\rightleftharpoons}}C11 &
      B12 \underset{{}}{\overset{{}}{\rightleftharpoons}}C12 \\
B13 \underset{{}}{\overset{{}}{\rightleftharpoons}}C13 &
  B14 \underset{{}}{\overset{{}}{\rightleftharpoons}}C14 &
    B15 \underset{{}}{\overset{{}}{\rightleftharpoons}}C15 \\
    B16 \underset{{}}{\overset{{}}{\rightleftharpoons}}C16 &
      B17 \underset{{}}{\overset{{}}{\rightleftharpoons}}C17 &
      B18 \underset{{}}{\overset{{}}{\rightleftharpoons}}C18 \\
      B19\underset{{}}{\overset{{}}{\rightleftharpoons}}C19 &
   B20 \underset{{}}{\overset{{}}{\rightleftharpoons}}C20 &
    B21 \underset{{}}{\overset{{}}{\rightleftharpoons}}C21 \\
     B22 \underset{{}}{\overset{{}}{\rightleftharpoons}}C22 &
     B23 \underset{{}}{\overset{{}}{\rightleftharpoons}}C23 &
       B24\underset{{}}{\overset{{}}{\rightleftharpoons}}C24 \\
B25 \underset{{}}{\overset{{}}{\rightleftharpoons}}C25 &
    B26 \underset{{}}{\overset{{}}{\rightleftharpoons}}C26 &
     B27 \underset{{}}{\overset{{}}{\rightleftharpoons}}C27 \\

\end{array}
\]

\begin{table}[!h]
\tabcolsep 5mm
\caption{parameters in MAPK cascade on scaffold}
\begin{center}
 \begin{tabular}{|p{1.2cm}|p{0.6cm}|p{0.6cm}|p{0.6cm}|p{0.6cm}|p{1.8cm}|p{6cm}|}
  \hline
Protein& on ($nM^{-1} $ min$^{-1}$)&off ($min^{-1}$)& **on
($nM^{-1}$ min$^{-1}$)& **off ($min^{-1}$)& phosphorylation ($
min^{-1})$& References and notes\\\hline Ste11 (MAP KKK)& 1.2 & 10
&1 &10 &20& Ste11pp can rebind to the scafold (27)
\\\hline Ste7 (MAP KK)& 1.2 & 0.1 &-& 0.5; 20& 46 & Ste7pp in the
scaffold undergoes hyper-phosphorylation by Fus3pp, which
accelerates Ste7's dissociation from the scaffold\\\hline Fus3
(MAPK)& 1.2 &10 &-&200& 250 & $t_{1/2}=0.3 s$ for the recovery of
Fuspp in shmoo tips(10,57)\\\hline
 \end{tabular}
 \end{center}
 \end{table}

\begin{table}[!h]
\tabcolsep 5mm
\caption{parameters in downstream reactions}
\begin{center}
\begin{tabular}{|c|c|p{10cm}|}
  \hline
protein&value&References and Notes\\\hline
  k28 &$ 8 nMmin^{-1}$& \\\hline
  k29 &$ 40 nM$& \\\hline
  k30 &$ 0.01 min^{-1} $& \\\hline
  k31 &$ 0.5 nMmin^{-1}$& \\\hline
  k32 &$ 10 nM $&  \\\hline
  k33&$ 2 nMmin^{-1}$& \\\hline
  k34 &$ 7 nM $& \\\hline
  k35 &$ 0.002 min^{-1}$& \\\hline
  k36 &$ 8 nM^{-1}min^{-1} $& \\\hline
  k37 &$ 1 min^{-1}$& \\\hline
  k38 &$ 10 min^{-1}$& \\\hline
  k39 &$ 10 min^{-1}$& \\\hline
  k40 &$ 0.1 nM^{-1}min^{-1}$& \\\hline
  k41 &$ 0.01 min^{-1} $& \\\hline
  k42 &$0.01 nM^{-1}min^{-1}$& \\\hline
  k43 &$ 0.1 min^{-1}$& \\\hline
  k44 &$ 32 nMmin^{-1} $& fit the time course of Sst2 determined in experiment (8)\\\hline
  k45 &$ 1 nM$& the same as k44\\\hline
  k46 &$ 0.052 min^{-1}$& the sam as k44\\\hline
  k47 &$ 10 min^{-1}$& $t_{1/2}=4s$ for the recovery of Fus3in nucleus (57)\\\cline{1-2}
  k48 &$ 14 min^{-1}$& $[Fus3_{in}]/[Fus3_{out}]=1.4$ at $t=0$ (45)\\\hline
  k49 &$ 8 min^{-1}$&  $[Fus3_{in}]/[Fus3_{out}]=2$ after 60 min treatment of  \\\cline{1-2}
  k50 &$ 15 min^{-1}$&$\alpha$-factor (45) \\
  \hline
\end{tabular}
\end{center}
\end{table}

\begin{table}[!h]
\tabcolsep 5mm
\caption{initial amount of proteins}
\begin{center}
\begin{tabular}{|l|c|p{11cm}|}
       \hline
       Protein &  value(nM) & notes and references \\ \hline
       Ste2 & 1000 & 8000 mole/cell (8) \\\hline
       Sst2 & 250 & 2000 mole/cell (8) \\\hline
       G protein & 1000 & 8000 mole/cell (8)\\\hline
       Ste20 & 1000 &  \\\hline
       Ste11 & 200 & less than 2000 mole/cell(1) \\\hline
       Fus3Ste7 & 200 & Total amount of Ste7 should be less than 2000 mole/cell, and about 95\% of \\\cline{1-2}
       Ste7 & 10 &     Ste7 is in the form of Fus3Ste7 before pheromone induction (1) \\\hline
       $Fus3_{in}$& 700 & about 5000 mole/cell (1), and before pheromone   \\\cline{1-2}
       $Fus3_{out}$& 300 & induction, $[Fus3_{in}]/[Fus3_{out}]=1.4$ (45) \\\hline
       MAPKKK-P  &  100  &  \\\hline
       MAPKK-P  & 100 &  \\\hline
       MAPK-P{\footnotesize in}& 100 &  \\\hline
       MAPK-P{\footnotesize out}&100& \\\hline
       Ste5{\footnotesize in} & 125 & (1) \\\hline
       Ste5{\footnotesize out} & 0.1 &  \\\hline
        Cdc28& 100  &  \\\hline
     \end{tabular}
     \end{center}
 \end{table}
{\footnotesize
\begin{table*}[!h]
\tabcolsep 5mm
\caption{mutants}
\begin{center}
\begin{tabular}{|c|p{3cm}|p{3cm}|p{3cm}|c|}
  \hline
  No. & Mutant & Behavior& Implementation & Reference\\\hline
  1 & $Ste2^{300\triangle}$ (endocytosis of receptor Ste2 is impaired) & G protein activation levels up & k3=0.08 & Ref.12 \\\hline
  2 & treated with cycloheximide(synthesis inhibitor) & G protein cycle is closed down &k31=k32=k44= k4=k9=0& Ref.12\\\hline
  3 & SST2$\triangle$ & super-sensitivity upon pheromone induction & k44=0,  $[Sst2]_{t=0}$=0 nM  & Ref.8\\\hline
  4 & 2 $\times$ SST2 & response upon pheromone reduces & k44=64, $[Sst2]_{t=0}$=500 nM & Ref.8\\\hline
  5 & 2 $\times$ G$\beta\gamma$ & super-sensitivity to pheromone induction & $[G\beta\gamma]_{t=0}=1000 nM$ & Ref.8\\\hline
  6 & negative feedbacks are cut down & activation of Fus3pp doesn't attenuate with time & k26=k27=k31= k44=0 &  \\\hline
  7 & dissociating speed for hyper-phosphorylated Ste7pp is slowed down &activation of Fus3pp doesn't attenuate with time & $**off_{KK}'=0.5 $& \\\hline
  8 & dissociating speed for normal Ste7pp is enhanced & activation of Fus3pp decreases & $**off_{KK}=20$ & \\\hline
  9 & shuttle of the scaffold Ste5 is cut down & super-sensitivity to pheromone induction & k22=k23=0 &\\\hline
  10 & Ste11 is continuously activated & Ste7pp is activated while Fus3pp is repressed down & $[Ste11pp]_{t=0}=200 nM, [Ste11]_{t=0}=0, [MAPKKK-P]_{totoal}=0, [\alpha-factor]=0$ & \\\hline
  11 & Ste7 is continuously activated  & Fus3pp isn't activated & $[Ste7pp]_{t=0}=210 nM, [Ste7]_{t=0}=0, [MAPKK-P]_{totoal}=0, [\alpha-factor]=0$& Ref.46\\
  \hline
\end{tabular}
\end{center}
\end{table*}
}

\begin{table}[!h]
\tabcolsep 5mm
\begin{flushleft}
\scriptsize{
\textbf{\emph{ODE functions}}:\\
$\large\displaystyle\frac{d[Ste2]}{dt}=-k1[\alpha-factor][Ste2]+k2[Ste2_{active}]-k7[Ste2]+\frac{k4[Ste12_{active}]^{2}}{k5^{2}+[Ste12_{active}]^{2}}+k6$\\
$\large\displaystyle\frac{d[Ste2_{active}]}{dt}=k1[\alpha-factor][Ste2]-k2[Ste2_{active}]-k3[Ste2_{active}] $\\
$\large\displaystyle\frac{d[Sst2_{active}]}{dt}=\frac{k44[Ste12_{active}]^2}{k45^2+[Ste12_{active}]^2}-k46[Sst2_{active}]$\\
$\large\displaystyle\frac{d[G]}{dt}=-k8[Ste2_{active}][G]+k15[G_{\alpha}{d}][G_{\beta}{\gamma}]+\frac{k9[Ste12_{active}]^{2}}{k10^{2}+[Ste12_{active}]^{2}}-k12[G]+k11$\\  
$\large\displaystyle\frac{d[G_{\alpha}{t}]}{dt}=k8[Ste2_{active}][G]-k13[G_{\alpha}{t}]-k14[G_{\alpha}{t}][Sst2_{active}]$\\   
$\large\displaystyle\frac{d[G_{\alpha}{d}]}{dt}=k13[G_{\alpha}{t}]+k14[G_{\alpha}{t}][Sst2_{active}]-k15[G_{\alpha}{d}][G_{\beta}{\gamma}]$ \\  
$\large\displaystyle\frac{d[G_{\beta}{\gamma}]}{dt}=k8[Ste2_{active}][G]-k15[G_{\alpha}{d}][G_{\beta}{\gamma}]-k40[G_{\beta}{\gamma}][Far1pp_{out}]+k41[Far1pp_{out}G_{\beta}{\gamma}]20-k18[G_{\beta}{\gamma}][Ste20]+k19[G_{\beta}{\gamma}Ste20]$ \\  
$\large\displaystyle\frac{d[Ste20]}{dt}=-k18[G_{\beta}{\gamma}][Ste20]+k19[G_{\beta}{\gamma}Ste20]$ \\ 
$\large\displaystyle\frac{d[G_{\beta}{\gamma} Ste20]}{dt}=k18[G_{\beta}{\gamma}][Ste20]-k19[G_{\beta}{\gamma} Ste20]-k16[G_{\beta}{\gamma}Ste20]B1+k17C1-k16[G_{\beta}{\gamma}Ste20]B2+k17C2-k16[G_{\beta}{\gamma}Ste20]B3+k17C3-k16[G_{\beta}{\gamma}Ste20]B4+k17C4-k16[G_{\beta}{\gamma}Ste20]B5+k17C5-k16[G_{\beta}{\gamma}Ste20]B6+k17C6-k16[G_{\beta}{\gamma}Ste20]B7+k17C7-k16[G_{\beta}{\gamma}Ste20]B8+k17C8-k16[G_{\beta}{\gamma}Ste20]B9+k17C9-k16[G_{\beta}{\gamma}Ste20]B10+k17C10-k16[G_{\beta}{\gamma}Ste20]B11+k17C11-k16[G_{\beta}{\gamma}Ste20]B12+k17C12-k16[G_{\beta}{\gamma}Ste20]B13+k17C13-k16[G_{\beta}{\gamma}Ste20]B14+k17C14-k16[G_{\beta}{\gamma}Ste20]B15+k17C15-k16[G_{\beta}{\gamma}Ste20]B16+k17C16-k16[G_{\beta}{\gamma}Ste20]B17+k17C17-k16[G_{\beta}{\gamma}Ste20]B18+k17C18-k16[G_{\beta}{\gamma}Ste20]B19+k17C19-k16[G_{\beta}{\gamma}Ste20]B20+k17C20-k16[G_{\beta}{\gamma}Ste20]B21+k17C21-k16[G_{\beta}{\gamma}Ste20]B22+k17C22-k16[G_{\beta}{\gamma}Ste20]B23+k17C23-k16[G_{\beta}{\gamma}Ste20]B24+k17C24-k16[G_{\beta}{\gamma}Ste20]B25+k17C25-k16[G_{\beta}{\gamma}Ste20]B26+k17C26-k16[G_{\beta}{\gamma}Ste20]B27+k17C27$\\
$\large\displaystyle\frac{d[Ste11]}{dt}=p1_{KKK}[Ste11pMAPKKK-P]+off_{KKK}(C2+C8+C11+C12+C15+C20+C22+C23+C26+B2+B8+B11+B12+B15+B20+B22+B23+B26)-on_{KKK}[Ste11](C4+C6+C7+C1+C16+C17+C18+C19+C5+B4+B6+B7+B1+B16+B17+B18+B19+B5)-k26[Ste11][Fus3pp_{out}]$\\
$\large\displaystyle\frac{d[Ste11p]}{dt}=-a1_{KKK}[Ste11p]([MAPKKK-P]_0-[Ste11pMAPKKK-P]-[Ste11ppMAPKKK-P])+d1_{KKK}[Ste11pMAPKKK-P]+p2_{KKK}[Ste11ppMAPKKK-P]$\\
$\large\displaystyle\frac{d[Ste11pMAPKKK-P]}{dt}=a1_{KKK}[Ste11p]([MAPKKK-P]_0-[Ste11pMAPKKK-P]-[Ste11ppMAPKKK-P])-(d1_{KKK}+p1_{KKK})[Ste11pMAPKKK-P]$\\  
$\large\displaystyle\frac{d[Ste11pp]}{dt}=-a2_{KKK}[Ste11pp]([MAPKKK-P]_0-[Ste11pMAPKKK-P]-[Ste11ppMAPKKK-P])+d2_{KKK}[Ste11ppMAPKKK-P]-a3_{KK}[Ste11pp][Ste7]+(d3_{KK}+p3_{KK})[Ste11ppSte7]-a4_{KK}[Ste11pp][Ste7p]+(d4_{KK}+p4_{KK})[Ste11ppSte7p]+**off_{KKK}(C3+C10+C9+C13+C14+C21+C24+C25+C27+B3+B10+B9+B13+B14+B21+B24+B25+B27) -**on_{KKK}[Ste11pp](C1+C4+C5+C6+C7+C16+C17+C18+C19+B1+B4+B5+B6+B7+B16+B17+B18+B19)$\\
$\large\displaystyle\frac{d[Ste11ppMAPKKK-P]}{dt}=a2_{KKK}[Ste11pp]([MAPKKK-P]_0-[Ste11pMAPKKK-P]-[Ste11ppMAPKKK-P])-(d2_{KKK}+p2_{KKK})[Ste11ppMAPKKK-P]$\\
$\large\displaystyle\frac{d[Ste7]}{dt}=-a3_{KK}[Ste7][Ste11pp]+d3_{KK}[Ste11ppSte7]+p1_{KK}[Ste7pMAPKK-P]+off_{KK}(C4+C8+C9+C16+C19+C20+C21+C23+C25+B4+B8+B9+B16+B19+B20+B21+B23+B25)-on_{KK}[Ste7](C1+C2+C3+C6+C7+C12+C13+C14+C15+B1+B2+B3+B6+B7+B12+B13+B14+B15)-k24[Ste7][Fus3_{out}]+k25[Fus3_{out}Ste7] $\\
$\large\displaystyle\frac{d[Ste7Ste11pp]}{dt}=a3_{KK}[Ste11pp][Ste7]-(d3_{KK}+p3_{KK})[Ste11ppSte7]$\\
$\large\displaystyle\frac{d[Ste7p]}{dt}=-a1_{KK}([MAPKK-P]_0-[Ste7pMAPKK-P]-[Ste7ppMAPKK-P])[Ste7p]+d1_{KK}[Ste7pMAPKK-P]+p3_{KK}[Ste11ppSte7]-a4_{KK}[Ste7p][Ste11pp]+d4_{KK}[Ste11ppSte7p]+p2_{KK}[Ste7ppMAPKK-P]$\\
$\large\displaystyle\frac{d[Ste7pMAPKK-P]}{dt}=a1_{KK}[Ste7p]([MAPKK-P]_0-[Ste7pMAPKK-P]-[Ste7ppMAPKK-P])-(d1_{KK}+p1_{KK})[Ste7pMAPKK-P]$ \\ 
$\large\displaystyle\frac{d[Ste7pSte11pp]}{dt}=a4_{KK}[Ste11pp][Ste7p]-(d4_{KK}+p4_{KK})[Ste11ppSte7p]$ \\ 
$\large\displaystyle\frac{d[Ste7pp]}{dt}=-a2_{KK}[Ste7pp]([MAPKK-P]_0-[Ste7pMAPKK-P]-[Ste7ppMAPKK-P])+d2_{KK}[Ste7ppMAPKK-P]+p4_{KK}[Ste11ppSte7p]-a3_{K}[Ste7pp][Fus3_{out}]+(d3_{K}+p3_{K})[Ste7ppFus3_{out}]-a4_{K}[Ste7pp][Fus3p_{out}]+(d4_{K}+p4_{K})[Ste7ppFus3p_{out}]+**off_{KK}(C5+C10+C11+C17+C22+C24+B5+B10+B11+B17+B22+B24)+**off_{KK}'(C18+C26+C27+B18+B26+B27)-k27[Ste7pp]$\\
$\large\displaystyle\frac{d[Ste7ppMAPKK-P]}{dt}=a2_{KK}[Ste7pp]([MAPKK-P]_0-[Ste7pMAPKK-P]-[Ste7ppMAPKK-P])-(d2_{KK}+p2_{KK})[Ste7ppMAPKK-P]$\\ 
$\large\displaystyle\frac{d[Fus3_{out}]}{dt}=-a3_{K}[Ste7pp][Fus3_{out}]+d3_{K}[Ste7ppFus3_{out}]+p1_{K}[Fus3p_{out}MAPK-P_{out}]+off_{K}(C6+C12+C13+C16+C17+C20+C21+C22+C24+B6+B12+B13+B16+B17+B20+B21+B22+B24)-on_{K}[Fus3_{out}](C1+C2+C3+C4+C5+C8+C9+C10+C11+B1+B2+B3+B4+B5+B8+B9+B10+B11)-k24[Ste7][Fus3_{out}]+k25[Fus3_{out}Ste7]+k47[Fus3_{in}]-k48[Fus3_{out}]+\frac{k32[Ste12_{active}]^{2}}{k5^{2}+[Ste12_{active}]^{2}}$\\
$\large\displaystyle\frac{d[Fus3_{out}Ste11pp]}{dt}=a3_{K}[Ste7pp][Fus3_{out}]-(d3_{K}+p3_{K})[Ste7ppFus3_{out}]$\\
$\large\displaystyle\frac{d[Fus3p_{out}]}{dt}=-a1_{K}[Fus3p_{out}][MAPK-P_{out}]+d1_{K}[Fus3p_{out}MAPK-P_{out}]+p3_{K}[Ste7ppFus3_{out}]-a4_{K}[Ste7pp][Fus3p_{out}]+d4_{K}[Ste7ppFus3p_{out}]+p2_{K}[Fus3pp_{out}MAPK-P_{out}]$\\
$\large\displaystyle\frac{d[Fus3p_{out}MAPK-P_{out}]}{dt}=a1_{K}[Fus3p_{out}][MAPK-P_{out}]-(d1_{K}+p1_{K})[Fus3p_{out}MAPK-P_{out}]$\\  
$\large\displaystyle\frac{d[Fus3p_{out}STe7pp]}{dt}=a4_{K}[Ste7pp][Fus3p_{out}]-(d4_{K}+p4_{K})[Ste7ppFus3p_{out}]$\\
$\large\displaystyle\frac{d[Fus3pp_{out}]}{dt}=-a2_{K}[Fus3pp_{out}][MAPK-P_{out}]+d2_{K}[Fus3pp_{out}MAPK-P_{out}]+p4_{K}[Ste7ppFus3p_{out}]+**off_{K}(C7+C14+C15+C18+C19+C23+C25+C26+C27+B7+B14+B15+B18+B19+B23+B25+B26+B27)+k49[Fus3pp_{in}]-k50[Fus3pp_{out}]$\\
$\large\displaystyle\frac{d[Fus3pp_{out}MAPK-P_{out}]}{dt}=a2_{K}[Fus3pp_{out}][MAPK-P_{out}]-(d2_{K}+p2_{K})[Fus3pp_{out}MAPK-P_{out}]$\\ 
$\large\displaystyle\frac{d[MAPK-P_{out}]}{dt}=-a1_{K}[Fus3p_{out}][MAPK-P_{out}]+(d1_{K}+p1_{K})[Fus3p_{out}MAPK-P_{out}]-a2_{K}[Fus3pp_{out}][MAPK-P_{out}]+(p2_{K}+d2_{K})[Fus3pp_{out}MAPK-P_{out}]+\frac{k31[Ste12_{active}]^{2}}{k5^{2}+[Ste12_{active}]^{2}}$ \\ 
$\large\displaystyle\frac{d[Fus3_{out}Ste7]}{dt}=k24[Ste7][Fus3_{out}]-k25[Fus3_{out}Ste7]$\\

}
\end{flushleft}
\end{table}
\begin{table}[!h]
\tabcolsep 5mm
\begin{flushleft}
\scriptsize{
$\large\displaystyle\frac{dB1}{dt}=-B1(on_{KKK}[Ste11]+**on_{KKK}[Ste11pp]+on_{KK}[Ste7]+on_{K}[Fus3_{out}])+off_{KKK}B2+**off_{KKK}B3+off_{KK}B4+**off_{KK}B5+off_{K}B6+**off_{K}B7-k16[G_{\beta}{\gamma}Ste20]B1+k17C1+k22[Ste5_{in}]-k23B1$\\
$\large\displaystyle\frac{dB2}{dt}=on_{KKK}B1[Ste11]-off_{KKK}B2-on_{KK}[Ste7]B2-on_{K}[Fus3_{out}]B2+off_{KK}B8+**off_{KK}B11+off_{K}B12+**off_{K}B15-k16[G_{\beta}{\gamma}Ste20]B2+k17C2$\\
$\large\displaystyle\frac{dB3}{dt}=**on_{KKK}[Ste11pp]B1-on_{KK}[Ste7]B3-on_{K}[Fus3_{out}]B3+off_{KK}B9+**off_{KK}B10+off_{K}B13+**off_{K}B14-**off_{KKK}B3-k16[G_{\beta}{\gamma}Ste20]B3+k17C3$\\
$\large\displaystyle\frac{dB4}{dt}=on_{KK}B1[Ste7]-off_{KK}B4-on_{KKK}[Ste11]B4-**on_{KKK}B4[Ste11pp]-on_{K}[Fus3_{out}]B4+off_{KKK}B8+**off_{KKK}B9+off_{K}B16+**off_{K}B19-k16[G_{\beta}{\gamma}Ste20]B4+k17C4$\\
$\large\displaystyle\frac{dB5}{dt}=-**on_{KKK}[Ste11pp]B5-on_{KKK}[Ste11]B5-on_{K}[Fus3_{out}]B5+**off_{KKK}B10+off_{KKK}B11+off_{K}B17+**off_{K}B18-**off_{KK}B5-k16[G_{\beta}{\gamma}Ste20]B5+k17C5$\\
$\large\displaystyle\frac{dB6}{dt}=-**on_{KKK}[Ste11pp]B6+on_{K}B1[Fus3_{out}]-off_{K}B6-on_{KKK}[Ste11]B6-on_{KK}[Ste7]B6+off_{KKK}B12+**off_{KKK}B13+off_{KK}B16+**off_{KK}B17-k16[G_{\beta}{\gamma}Ste20]B6+k17C6$\\
$\large\displaystyle\frac{dB7}{dt}=-**on_{KKK}[Ste11pp]B7-on_{KKK}[Ste11]B7-on_{KK}[Ste7]B7+**off_{KKK}B14+off_{KKK}B15+**off_{KK}'B18+off_{KK}B19-**off_{K}B7-k16[G_{\beta}{\gamma}Ste20]B7+k17C7$\\
$\large\displaystyle\frac{dB8}{dt}=-on_{K}[Fus3_{out}]B8-(off_{KKK}+off_{KK})B8+on_{KKK}[Ste11]B4+on_{KK}[Ste7]B2+off_{K}B20+**off_{K}B23-k16[G_{\beta}{\gamma}Ste20]B8+k17C8$\\
$\large\displaystyle\frac{dB9}{dt}=**on_{KKK}B4[Ste11pp]-p_{KK}B9-(**off_{KKK}+off_{KK})B9-on_{K}[Fus3_{out}]B9+on_{KK}[Ste7]B3+off_{K}B21+**off_{K}B25-k16[G_{\beta}{\gamma}Ste20]B9+k17C9$\\
$\large\displaystyle\frac{dB10}{dt}=**on_{KKK}[Ste11pp]B5-(**off_{KKK}+**off_{KK})B10-on_{K}[Fus3_{out}]B10+off_{K}B24+**off_{K}B27+p_{KK}B9-k16[G_{\beta}{\gamma}Ste20]B10+k17C10$\\
$\large\displaystyle\frac{dB11}{dt}=-(off_{KKK}+**off_{KK})B11-on_{K}[Fus3_{out}]B11+on_{KKK}[Ste11]B5+off_{K}B22+**off_{K}B26-k16[G_{\beta}{\gamma}Ste20]B11+k17C11$\\
$\large\displaystyle\frac{dB12}{dt}=-on_{KK}[Ste7]B12-(off_{KKK}+off_{K})B12+on_{KKK}[Ste11]B6+on_{K}[Fus3_{out}]B2+off_{KK}B20+**off_{KK}B22-k16[G_{\beta}{\gamma}Ste20]B12+k17C12$\\
$\large\displaystyle\frac{dB13}{dt}=**on_{KKK}[Ste11pp]B6-(**off_{KKK}+off_{K})B13-on_{KK}[Ste7]B13+on_{K}[Fus3_{out}]B3+off_{KK}B21+**off_{KK}B24-k16[G_{\beta}{\gamma}Ste20]B13+k17C13$\\
$\large\displaystyle\frac{dB14}{dt}=**on_{KKK}[Ste11pp]B7-(**off_{KKK}+**off_{K})B14-on_{KK}[Ste7]B14+off_{KK}B25+**off_{KK}'B27-k16[G_{\beta}{\gamma}Ste20]B14+k17C14$\\
$\large\displaystyle\frac{dB15}{dt}=-(off_{KKK}+**off_{K})B15-on_{KK}[Ste7]B15+on_{KKK}[Ste11]B7+off_{KK}B23+**off_{KK}'B26-k16[G_{\beta}{\gamma}Ste20]B15+k17C15$\\
$\large\displaystyle\frac{dB16}{dt}=-**on_{KKK}[Ste11pp]B16-on_{KKK}[Ste11]B16-(off_{KK}+off_{K})B16+on_{KK}[Ste7]B6+on_{K}[Fus3_{out}]B4+off_{KKK}B20+**off_{KKK}B21-k16[G_{\beta}{\gamma}Ste20]B16+k17C16$\\
$\large\displaystyle\frac{dB17}{dt}=-**on_{KKK}[Ste11pp]B17-p_{K}B17-(**off_{KK}+off_{K})B17-on_{KKK}[Ste11]B17+on_{K}[Fus3_{out}]B5+off_{KKK}B22+**off_{KKK}B24-k16[G_{\beta}{\gamma}Ste20]B17+k17C17$\\
$\large\displaystyle\frac{dB18}{dt}=-**on_{KKK}[Ste11pp]B18-(**off_{KK}'+**off_{K})B18-on_{KKK}[Ste11]B18+off_{KKK}B26+**off_{KKK}B27+p_{K}B17-k16[G_{\beta}{\gamma}Ste20]B18+k17C18$\\
$\large\displaystyle\frac{dB19}{dt}=-**on_{KKK}[Ste11pp]B19-(off_{KK}+**off_{K})B19-on_{KKK}[Ste11]B19+on_{KK}[Ste7]B7+off_{KKK}B23+**off_{KKK}B25-k16[G_{\beta}{\gamma}Ste20]B19+k17C19$\\
$\large\displaystyle\frac{dB20}{dt}=-(off_{KKK}+off_{KK}+off_{K})B20+on_{KKK}[Ste11]B16+on_{KK}[Ste7]B12+on_{K}[Fus3_{out}]B8-k16[G_{\beta}{\gamma}Ste20]B20+k17C20$\\
$\large\displaystyle\frac{dB21}{dt}=**on_{KKK}[Ste11pp]B16-(**off_{KKK}+off_{KK}+off_{K})B21-p_{KK}B21+on_{KK}[Ste7]B13+on_{K}[Fus3_{out}]B9-k16[G_{\beta}{\gamma}Ste20]B21+k17C21$\\
$\large\displaystyle\frac{dB22}{dt}=-(off_{KKK}+**off_{KK}+off_{K})B22-(p_{K})B22+on_{KKK}[Ste11]B17+on_{K}[Fus3_{out}]B11-k16[G_{\beta}{\gamma}Ste20]B22+k17C22$\\
$\large\displaystyle\frac{dB23}{dt}=-(off_{KKK}+off_{KK}+**off_{K})B23+on_{KKK}[Ste11]B19+on_{KK}[Ste7]B15-k16[G_{\beta}{\gamma}Ste20]B23+k17C23$\\
$\large\displaystyle\frac{dB24}{dt}=**on_{KKK}[Ste11pp]B17-(**off_{KKK}+**off_{KK}+off_{K})B24-p_{K}B24+on_{K}[Fus3_{out}]B10+p_{KK}B21-k16[G_{\beta}{\gamma}Ste20]B24+k17C24$\\
$\large\displaystyle\frac{dB25}{dt}=**on_{KKK}[Ste11pp]B19-(**off_{KKK}+off_{KK}+**off_{K})B25-p_{KK}B25+on_{KK}[Ste7]B14-k16[G_{\beta}{\gamma}Ste20]B25+k17C25$\\
$\large\displaystyle\frac{dB26}{dt}=-(off_{KKK}+**off_{KK}'+**off_{K})B26+on_{KKK}[Ste11]B18+p_{K}B22-k16[G_{\beta}{\gamma}Ste20]B26+k17C26$\\
$\large\displaystyle\frac{dB27}{dt}=**on_{KKK}[Ste11pp]B18-(**off_{KKK}+**off_{KK}'+**off_{K})B27+p_{KK}B25+p_{K}B24-k16[G_{\beta}{\gamma}Ste20]B27+k17C27$ \\
}
\end{flushleft}
\end{table}
\begin{table}[!h]
\tabcolsep 5mm
\begin{flushleft}
\scriptsize{
$\large\displaystyle\frac{dC1}{dt}=-C1(on_{KKK}[Ste11]+**on_{KKK}[Ste11pp]+on_{KK}[Ste7]+on_{K}[Fus3_{out}])+off_{KKK}C2+**off_{KKK}C3+off_{KK}C4+**off_{KK}C5+off_{K}C6+**off_{K}C7+k16[G_{\beta}{\gamma}Ste20]B1-k17C1$\\
$\large\displaystyle\frac{dC2}{dt}=on_{KKK}C1[Ste11]-off_{KKK}C2-p_{KKK}C2-on_{KK}[Ste7]C2-on_{K}[Fus3_{out}]C2+off_{KK}C8+**off_{KK}C11+off_{K}C12+**off_{K}C15+k16[G_{\beta}{\gamma}Ste20]B2-k17C2$\\
$\large\displaystyle\frac{dC3}{dt}=**on_{KKK}[Ste11pp]C1+p_{KKK}C2-on_{KK}[Ste7]C3-on_{K}[Fus3_{out}]C3+off_{KK}C9+**off_{KK}C10+off_{K}C13+**off_{K}C14-**off_{KKK}C3+k16[G_{\beta}{\gamma}Ste20]B3-k17C3$\\
$\large\displaystyle\frac{dC4}{dt}=on_{KK}C1[Ste7]-off_{KK}C4-on_{KKK}[Ste11]C4-**on_{KKK}C4[Ste11pp]-on_{K}[Fus3_{out}]C4+off_{KKK}C8+**off_{KKK}C9+off_{K}C16+**off_{K}C19+k16[G_{\beta}{\gamma}Ste20]B4-k17C4$\\
$\large\displaystyle\frac{dC5}{dt}=-**on_{KKK}[Ste11pp]C5-on_{KKK}[Ste11]C5-on_{K}[Fus3_{out}]C5+**off_{KKK}C10+off_{KKK}C11+off_{K}C17+**off_{K}C18-**off_{KK}C5+k16[G_{\beta}{\gamma}Ste20]B5-k17C5$\\
$\large\displaystyle\frac{dC6}{dt}=-**on_{KKK}[Ste11pp]C6+on_{K}C1[Fus3_{out}]-off_{K}C6-on_{KKK}[Ste11]C6-on_{KK}[Ste7]C6+off_{KKK}C12+**off_{KKK}C13+off_{KK}C16+**off_{KK}C17+k16[G_{\beta}{\gamma}Ste20]B6-k17C6$\\
$\large\displaystyle\frac{dC7}{dt}=-**on_{KKK}[Ste11pp]C7-on_{KKK}[Ste11]C7-on_{KK}[Ste7]C7+**off_{KKK}C14+off_{KKK}C15+**off_{KK}'C18+off_{KK}C19-**off_{K}C7+k16[G_{\beta}{\gamma}Ste20]B7-k17C7$\\
$\large\displaystyle\frac{dC8}{dt}=-p_{KKK}C8-on_{K}[Fus3_{out}]C8-(off_{KKK}+off_{KK})C8+on_{KKK}[Ste11]C4+on_{KK}[Ste7]C2+off_{K}C20+**off_{K}C23+k16[G_{\beta}{\gamma}Ste20]B8-k17C8$\\
$\large\displaystyle\frac{dC9}{dt}=**on_{KKK}C4[Ste11pp]-p_{KK}C9-(**off_{KKK}+off_{KK})C9-on_{K}[Fus3_{out}]C9+on_{KK}[Ste7]C3+off_{K}C21+**off_{K}C25+p_{KKK}C8+k16[G_{\beta}{\gamma}Ste20]B9-k17C9$\\
$\large\displaystyle\frac{dC10}{dt}=**on_{KKK}[Ste11pp]C5-(**off_{KKK}+**off_{KK})C10-on_{K}[Fus3_{out}]C10+off_{K}C24+**off_{K}C27+p_{KKK}C11+p_{KK}C9+k16[G_{\beta}{\gamma}Ste20]B10-k17C10$\\
$\large\displaystyle\frac{dC11}{dt}=-p_{KKK}C11-(off_{KKK}+**off_{KK})C11-on_{K}[Fus3_{out}]C11+on_{KKK}[Ste11]C5+off_{K}C22+**off_{K}C26+k16[G_{\beta}{\gamma}Ste20]B11-k17C11$\\
$\large\displaystyle\frac{dC12}{dt}=-p_{KKK}C12-on_{KK}[Ste7]C12-(off_{KKK}+off_{K})C12+on_{KKK}[Ste11]C6+on_{K}[Fus3_{out}]C2+off_{KK}C20+**off_{KK}C22+k16[G_{\beta}{\gamma}Ste20]B12-k17C12$\\
$\large\displaystyle\frac{dC13}{dt}=**on_{KKK}[Ste11pp]C6-(**off_{KKK}+off_{K})C13-on_{KK}[Ste7]C13+on_{K}[Fus3_{out}]C3+off_{KK}C21+**off_{KK}C24+p_{KKK}C12+k16[G_{\beta}{\gamma}Ste20]B13-k17C13$\\
$\large\displaystyle\frac{dC14}{dt}=**on_{KKK}[Ste11pp]C7-(**off_{KKK}+**off_{K})C14-on_{KK}[Ste7]C14+off_{KK}C25+**off_{KK}'C27+p_{KKK}C15+k16[G_{\beta}{\gamma}Ste20]B14-k17C14$\\
$\large\displaystyle\frac{dC15}{dt}=-p_{KKK}C15-(off_{KKK}+**off_{K})C15-on_{KK}[Ste7]C15+on_{KKK}[Ste11]C7+off_{KK}C23+**off_{KK}'C26+k16[G_{\beta}{\gamma}Ste20]B15-k17C15$\\
$\large\displaystyle\frac{dC16}{dt}=-**on_{KKK}[Ste11pp]C16-on_{KKK}[Ste11]C16-(off_{KK}+off_{K})C16+on_{KK}[Ste7]C6+on_{K}[Fus3_{out}]C4+off_{KKK}C20+**off_{KKK}C21+k16[G_{\beta}{\gamma}Ste20]B16-k17C16$\\
$\large\displaystyle\frac{dC17}{dt}=-**on_{KKK}[Ste11pp]C17-p_{K}C17-(**off_{KK}+off_{K})C17-on_{KKK}[Ste11]C17+on_{K}[Fus3_{out}]C5+off_{KKK}C22+**off_{KKK}C24+k16[G_{\beta}{\gamma}Ste20]B17-k17C17$\\
$\large\displaystyle\frac{dC18}{dt}=-**on_{KKK}[Ste11pp]C18-(**off_{KK}'+**off_{K})C18-on_{KKK}[Ste11]C18+off_{KKK}C26+**off_{KKK}C27+p_{K}C17+k16[G_{\beta}{\gamma}Ste20]B18-k17C18$\\
$\large\displaystyle\frac{dC19}{dt}=-**on_{KKK}[Ste11pp]C19-(off_{KK}+**off_{K})C19-on_{KKK}[Ste11]C19+on_{KK}[Ste7]C7+off_{KKK}C23+**off_{KKK}C25+k16[G_{\beta}{\gamma}Ste20]B19-k17C19$\\
$\large\displaystyle\frac{dC20}{dt}=-(off_{KKK}+off_{KK}+off_{K})C20-p_{KKK}C20+on_{KKK}[Ste11]C16+on_{KK}[Ste7]C12+on_{K}[Fus3_{out}]C8+k16[G_{\beta}{\gamma}Ste20]B20-k17C20$\\
$\large\displaystyle\frac{dCB21}{dt}=**on_{KKK}[Ste11pp]C16-(**off_{KKK}+off_{KK}+off_{K})C21-p_{KK}C21+on_{KK}[Ste7]C13+on_{K}[Fus3_{out}]C9+p_{KKK}C20+k16[G_{\beta}{\gamma}Ste20]B21-k17C21$\\
$\large\displaystyle\frac{dC22}{dt}=-(off_{KKK}+**off_{KK}+off_{K})C22-(p_{KKK}+p_{K})C22+on_{KKK}[Ste11]C17+on_{K}[Fus3_{out}]C11+k16[G_{\beta}{\gamma}Ste20]B22-k17C22$\\
$\large\displaystyle\frac{dC23}{dt}=-(off_{KKK}+off_{KK}+**off_{K})C23-p_{KKK}C23+on_{KKK}[Ste11]C19+on_{KK}[Ste7]C15+k16[G_{\beta}{\gamma}Ste20]B23-k17C23$\\
$\large\displaystyle\frac{dC24}{dt}=**on_{KKK}[Ste11pp]C17-(**off_{KKK}+**off_{KK}+off_{K})C24-p_{K}C24+on_{K}[Fus3_{out}]C10+p_{KKK}C22+p_{KK}C21+k16[G_{\beta}{\gamma}Ste20]B24-k17C24$\\
$\large\displaystyle\frac{dC25}{dt}=**on_{KKK}[Ste11pp]C19-(**off_{KKK}+off_{KK}+**off_{K})C25-p_{KK}C25+on_{KK}[Ste7]C14+p_{KKK}C23+k16[G_{\beta}{\gamma}Ste20]B25-k17C25$\\
$\large\displaystyle\frac{dC26}{dt}=-(off_{KKK}+**off_{KK}'+**off_{K})C26-p_{KKK}C26+on_{KKK}[Ste11]C18+p_{K}C22+k16[G_{\beta}{\gamma}Ste20]B26-k17C26$\\
$\large\displaystyle\frac{dC27}{dt}=**on_{KKK}[Ste11pp]C18-(**off_{KKK}+**off_{KK}'+**off_{K})C27+p_{KKK}C26+p_{KK}C25+p_{K}C24+k16[G_{\beta}{\gamma}Ste20]B27-k17C27$ \\
}
\end{flushleft}
\end{table}
\begin{table}[!h]
\tabcolsep 5mm
\begin{flushleft}
\scriptsize{
$\large\displaystyle\frac{d[Ste5_{in}]}{dt}= -k22[Ste5_{in}]+k23B1$\\
$\large\displaystyle\frac{d[Fus3_{in}]}{dt}=-k47[Fus3_{in}]+k48[Fus3_{out}]+p1_{K}[Fus3p_{in}MAPK-P_{in}]$ \\ 
$\large\displaystyle\frac{d[Fus3p_{in}]}{dt}=p2_{K}[Fus3pp_{in}MAPK-P_{in}]-a1_{K}[Fus3p_{in}][MAPK-P_{in}]+d1_{K}[Fus3p_{in}MAPK-P_{in}]$\\
$\large\displaystyle\frac{d[Fus3p_{in}MAPK-P_{in}]}{dt}=a1_{K}[Fus3p_{in}][MAPK-P_{in}]-d1_{K}[Fus3p_{in}MAPK-P_{in}]-p1_{K}[Fus3p_{in}MAPK-P_{in}]$\\
$\large\displaystyle\frac{d[Fus3pp_{in}]}{dt}=-k49[Fus3pp_{in}]+k50[Fus3pp_{out}]-a2_{K}[Fus3pp_{in}][MAPK-P_{in}]+d2_{K}[Fus3pp_{in}MAPK-P_{in}]$ \\
$\large\displaystyle\frac{d[Fus3pp_{in}MAPK-P_{in}]}{dt}=a2_{K}[Fus3pp_{in}][MAPK-P_{in}]-d2_{K}[Fus3pp_{in}MAPK-P_{in}]-p2_{K}[Fus3pp_{in}MAPK-P_{in}]$\\
$\large\displaystyle\frac{d[MAPK-P_{in}]}{dt}=-a1_{K}[Fus3p_{in}][MAPK-P_{in}]+(d1_{K}+p1_{K})[Fus3p_{in}MAPK-P_{in}]-a2_{K}[Fus3pp_{in}][MAPK-P_{in}]+(p2_{K}+d2_{K})[Fus3pp_{in}MAPK-P_{in}]$\\
$\large\displaystyle\frac{d[Ste12_{active}]}{dt}=\frac{k28[Fus3pp_{in}]}{k29+[Fus3pp_{in}]}-k30[Ste12_{active}]$\\  
$\large\displaystyle\frac{d[Far1]}{dt}=\frac{k33[Ste12_{active}]^{2}}{k34^{2}+[Ste12_{active}]^{2}}-k36[Far1][Fus3pp_{in}]-k35[Far1]+k37[Far1pp_{in}]$ \\ 
$\large\displaystyle\frac{d[Far1pp_{in}]}{dt}=k36[Far1][Fus3pp_{in}]-k37[Far1pp_{in}]-k38[Far1pp_{in}]+k39[Far1pp_{out}]-k43[Far1pp_{in}][Cdc28]+k42[Far1pp_{in}Cdc28]$ \\ 
$\large\displaystyle\frac{d[Cdc28]}{dt}=k42[Far1pp_{in}Cdc28]-k43[Far1pp_{in}][Cdc28]$\\  
$\large\displaystyle\frac{d[Far1pp_{in}Cdc28]}{dt}=-k42[Far1pp_{in}Cdc28]+k43[Far1pp_{in}][Cdc28]$ \\
$\large\displaystyle\frac{d[Far1pp_{out}]}{dt}=-k40[Far1pp_{out}][G_{\beta}{\gamma}]+k41[Far1pp_{out}G_{\beta}{\gamma}]+k38[Far1pp_{in}]-k39[Far1pp_{out}]$\\
$\large\displaystyle\frac{d[Far1pp_{out}G_{\beta}{\gamma}]}{dt}=k40[G_{\beta}{\gamma}][Far1pp_{out}]-k41[Far1pp_{out}G_{\beta}{\gamma}]$\\
}
\end{flushleft}
\end{table}